\documentclass[lettersize,journal]{IEEEtran}
\usepackage{amsmath,amsfonts}
\usepackage{algorithmic}
\usepackage{algorithm}
\usepackage{array}
\usepackage[caption=false,font=normalsize,labelfont=sf,textfont=sf]{subfig}
\usepackage{textcomp}
\usepackage{stfloats}
\usepackage{url}
\usepackage{verbatim}
\usepackage{graphicx}
\usepackage{cite}
\usepackage{booktabs}
\usepackage{amssymb}
\usepackage{bbm}
\usepackage{tabularx}
\usepackage{xcolor}
\usepackage[acronym,nomain]{glossaries}
\makeglossaries

\newacronym{IAB}{IAB}{Integrated Access and Backhaul}
\newacronym{GNN}{GNN}{Graph Neural Network}
\newacronym{GATv2}{GATv2}{Graph Attention Network v2}
\newacronym{GCN}{GCN}{Graph Convolutional Network}
\newacronym{MDP}{MDP}{Markov Decision Process}
\newacronym{PPO}{PPO}{Proximal Policy Optimization}
\newacronym{DQN}{DQN}{Deep Q-Network}
\newacronym{MIP}{MIP}{Mixed-Integer Programming}
\newacronym{DRL}{DRL}{Deep Reinforcement Learning}
\newacronym{PSO}{PSO}{Particle Swarm Optimization}
\newacronym{VNF}{VNF}{Virtual Network Function}
\newacronym{RAN}{RAN}{Radio Access Network}
\newacronym{O-RUs}{O-RUs}{Open Radio Units}
\newacronym{O-DUs}{O-DUs}{Open Distributed Units}
\newacronym{O-CUs}{O-CUs}{Open Central Units}
\newacronym{RL}{RL}{reinforcement learning}
\newacronym{UAV}{UAV}{Unmanned Aerial Vehicle}
\newacronym{IoT}{IoT}{Internet of Things}
\newacronym{MIMO}{MIMO}{Multiple-Input Multiple-Output}
\newacronym{eMBB}{eMBB}{enhanced Mobile Broadband}
\newacronym{QoS}{QoS}{Quality of Service}
\newacronym{SNR}{SNR}{Signal to Noise Ratio}
\newacronym{CSI}{CSI}{Channel State Information}
\newacronym{DT}{DT}{Digital Twin}
\newacronym{MT}{MT}{Mobile Termination}
\newacronym{DU}{DU}{Distributed Unit}
\newacronym{mmWave}{mmWave}{millimeter-wave}
\newacronym{GIS}{GIS}{Geographic Information System}
\newacronym{6G}{6G}{Sixth Generation}

\newacronym{FR2}{FR2}{Frequency Range 2}

\newacronym{3GPP}{3GPP}{3rd Generation Partnership Project}
\newacronym{NR}{NR}{New Radio}

\newacronym{O-RAN}{O-RAN}{Open Radio Access Network}

\newacronym{NSGA-II}{NSGA-II}{Non-dominated Sorting Genetic Algorithm II}
\newacronym{MOEA/D}{MOEA/D}{Multi-Objective Evolutionary Algorithm based on Decomposition}
\newacronym{SAC}{SAC}{Soft Actor-Critic}
\begin{document}

\title{Resilient Integrated Access and Backhaul Network Planning via Graph Neural Network Curriculum Learning}

\author{
Jie~Zhang,
Haochen Sun,
Mingzhang Zhou,
Paul~D.~Mitchell, 
Hamed~Ahmadi,
and~Qiang~Ni
\thanks{This work was supported in part by the UKRI EPSRC and the Department for Transport (DfT) via the project ``CRDT-Port: Developing a Crisis and Resilience Digital Twin for UK Port-centric Transport Systems'' under Grant UKRI858; in part by the U.K. Engineering and Physical Sciences Research Council (EPSRC), Hub for All Spectrum Communications (HASC), through the O-RAN Intelligent Adaptive Load Balancing and Efficiency in Highly Dense Deployments (ORLANDO) Project under Grant EP/X040569/1; and in part by EPSRC Impact Acceleration Accounts (IAA) through the Green Secure and Privacy Aware Wireless Networks for Sustainable Future Connected and Autonomous Systems project under Grant EP/X525856/1; and in part by the EPSRC CyberFocus project under Grant PBIAA-CF-003-QN-LU.}
\thanks{J. Zhang, Haochen Sun, M. Zhou, P. D. Mitchell, and H. Ahmadi are with the
School of Physics, Engineering and Technology (PET), University of York,
York YO10 5DD, United Kingdom
(e-mail: jie.zhang@york.ac.uk).}
\thanks{Q. Ni is with the School of Computing and Communications,
Lancaster University, Lancaster LA1 4YW, United Kingdom.}
\thanks{Copyright (c) 2026 IEEE. Personal use of this material is permitted. However, permission to use this material for any other purposes must be obtained from the IEEE by sending a request to pubs-permissions@ieee.org.}
}

\markboth{IEEE INTERNET OF THINGS JOURNAL}%
{Zhang \MakeLowercase{\textit{et al.}}: Resilience-Aware mmWave IAB Deployment Planning via Graph Neural Network Curriculum Learning}


\maketitle
\begin{abstract}
Integrated Access and Backhaul (IAB) is a key enabler for cost-effective millimeter-wave (mmWave) densification, yet practical deployments are strongly constrained by the spatial availability of fiber-connected IAB donors. Variations in donor topologies reshape both the initial coverage footprint and the feasible multi-hop backhaul expansion patterns, causing planning policies trained on a single topology to generalize poorly. This paper studies cross-topology generalization for resilience-aware IAB deployment under coverage and transport feasibility requirements. We formulate IAB deployment as a constrained Markov decision process that couples access coverage satisfaction with multi-hop transport feasibility, including hop-by-hop rate-budget degradation and a minimum parent-connectivity requirement for fault tolerance. To solve the resulting large discrete planning problem, we propose a graph-based actor--critic approach that uses a Graph Attention Network v2 (GATv2) encoder with Proximal Policy Optimization (PPO), and we develop a multi-stage curriculum learning strategy that progressively increases constraint difficulty and scenario diversity for stable training and transferable policies. In a 60\,GHz urban simulation over a $1000\,\mathrm{m}\times1000\,\mathrm{m}$ area with multiple donor layouts, the trained policy satisfies the minimum coverage requirement of 98\% with fewer deployed IAB nodes than learning and heuristic baselines, and maintains higher coverage under progressive backhaul failures. The results demonstrate that curriculum-trained graph-based reinforcement learning can improve deployment efficiency and resilience while transferring across heterogeneous IAB donor topologies without retraining.
\end{abstract}


\section{Introduction}
\label{sec:introduction_tnse}

The evolution toward dense 6G infrastructures is increasingly shaped by two coupled forces: the migration to high-frequency spectrum (e.g., \gls{mmWave}/\gls{FR2}) and the continued densification of access points in urban environments. Ultra-dense deployments reduce inter-site distances and enhance spatial reuse, but they also shift the practical bottleneck from the access link to the transport layer, which must sustain end-to-end service quality at scale \cite{kamel2016ultra,giordani2020toward}. In \gls{mmWave} bands, the transport layer is particularly fragile because link availability and capacity are strongly influenced by environmental geometry, blockage dynamics, and beam-management overhead \cite{Rappaport2013Access,Rangan2014JPROC,MacCartney2017,siles2015atmospheric}. Consequently, the network-wide performance of dense \gls{6G} systems is no longer determined solely by access coverage, but by the joint feasibility of multi-hop transport, interference-aware scheduling, and resource multiplexing under highly variable link conditions.

To address these transport-layer challenges in a cost-effective manner, \gls{3GPP} introduced \gls{IAB} as a unified framework that reuses \gls{NR} radio and protocol functions for both access and wireless backhaul, enabling multi-hop transport without ubiquitous fiber deployment \cite{TR38874,TS38300,TS38401}. Empirical evidence and system-level evaluations indicate that \gls{IAB} can accelerate \gls{mmWave} rollouts and extend coverage and capacity in areas where fixed backhaul is prohibitively expensive or operationally infeasible \cite{polese2020integrated,cudak2021integrated,ronkainen2020integrated}. Beyond its role as a standardized solution for cellular backhauling, \gls{IAB} exemplifies a broader class of hierarchical multi-hop wireless infrastructures.

Nevertheless, deploying \gls{IAB} networks introduces planning challenges that differ fundamentally from those of conventional cellular design. In traditional networks, the transport layer is typically assumed to be static and reliable, and planning therefore focuses primarily on site placement and access coverage. In \gls{IAB}, by contrast, the transport layer itself becomes a design variable. As a result, topology formation, routing, and scheduling are tightly coupled: half-duplex operation, interference, and traffic aggregation induce dependencies that propagate across tiers, such that even local decisions can affect end-to-end service across multiple hops. This coupling also renders the planning problem intrinsically difficult. Related results in multi-radio mesh networks show that joint optimization of routing and channel assignment is computationally challenging \cite{Alicherry2006MobiCom}, while classical throughput-scaling laws indicate that multi-hop wireless capacity can degrade rapidly as hop count increases, making transport a first-order constraint in dense deployments \cite{Gupta2000Capacity}.

\begin{figure}[!t]
\centering
\includegraphics[width=\columnwidth]{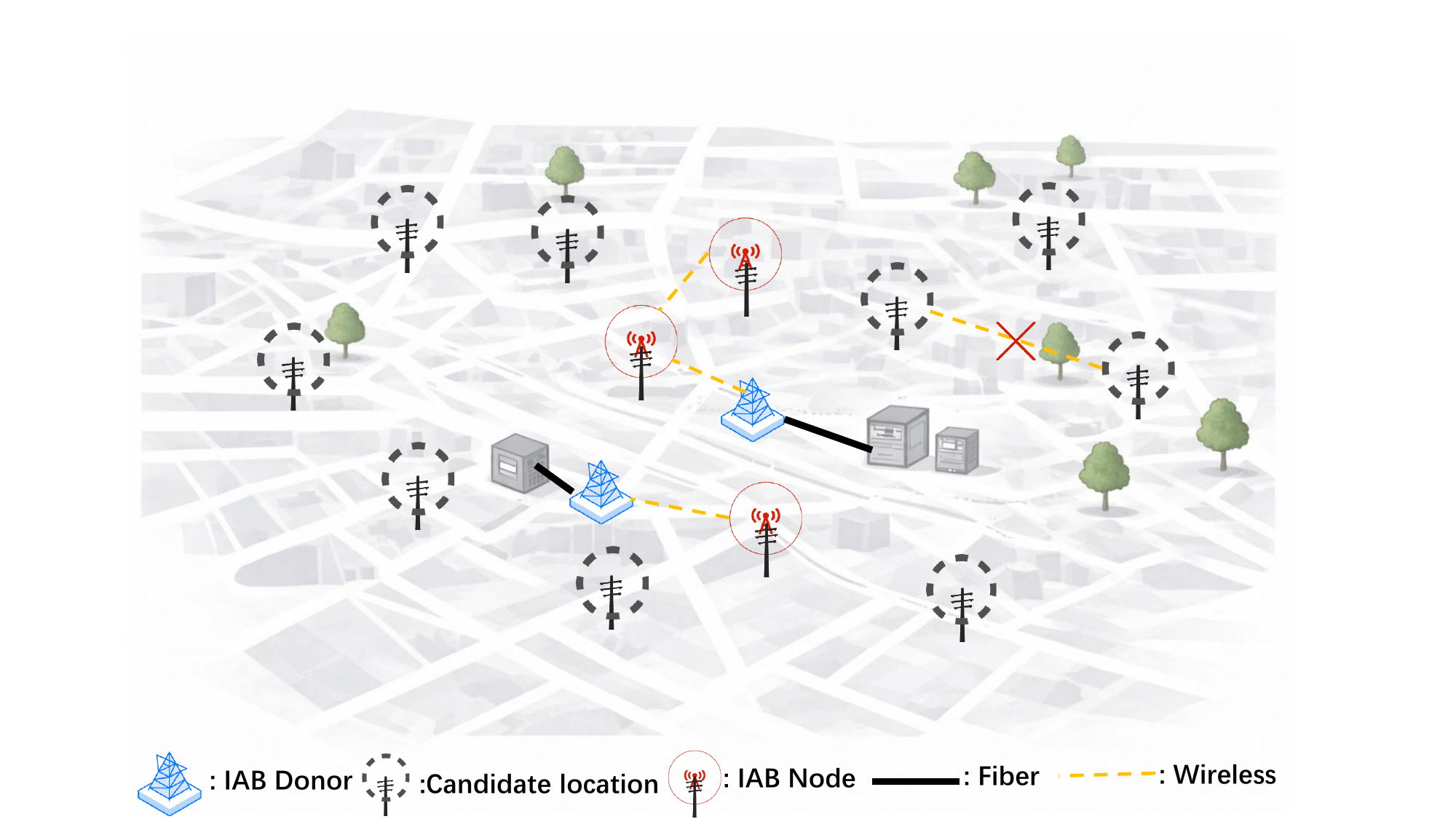}
\caption{Representative mmWave IAB deployment scenario showing fiber-connected IAB donors, candidate sites, deployed IAB nodes, and wireless backhaul links.}

\label{fig:scenario_overview}
\end{figure}

In addition to efficiency, dense \gls{mmWave} multi-hop transport imposes stringent resilience requirements. Wireless backhaul links may degrade or fail due to blockage, weather, or local environmental changes \cite{MacCartney2017,polese2020integrated}. A resilient communication network should maintain acceptable service under faults and recover efficiently following disruptions \cite{Sterbenz2010Resilience}. For \gls{IAB}, resilience is particularly challenging because \gls{IAB} donors must be fiber-connected to the core network, constraining donor availability to a location-dependent set of sites. Consequently, practical deployments begin from heterogeneous \gls{IAB} donor topologies—ranging from dispersed and near-symmetric layouts to strongly anisotropic or clustered configurations—depending on the fiber footprint and urban constraints \cite{polese2020integrated}. This heterogeneity affects the initial coverage footprint and, more importantly, reshapes feasible multi-hop expansion patterns, thereby determining how quickly transport resources become the limiting factor. Planning policies optimized for a single \gls{IAB} donor topology therefore often generalize poorly to other donor configurations, producing deployments that are structurally brittle and more vulnerable to \gls{mmWave} link degradations. Fig.~\ref{fig:scenario_overview} shows a representative mmWave IAB deployment scenario, where fiber-connected \gls{IAB} donors and deployed \gls{IAB} nodes expand coverage via multi-hop wireless backhaul; an example blocked backhaul link is also illustrated.

Accordingly, effective \gls{IAB} deployment requires a planning mechanism that can exploit the relational structure induced by the donor topology and the evolving multi-hop backhaul. \glspl{GNN} provide a natural means of representing deployments as graphs and modeling multi-hop dependencies through permutation-invariant message passing \cite{scarselli2008graph,velivckovic2017graph}. When combined with \gls{RL}, which supports sequential decision-making under uncertainty \cite{sutton2018reinforcement}, graph-based learning can yield planning policies that adapt to changing conditions while respecting topological constraints. This direction also aligns with data-driven \gls{RAN} control, including \gls{O-RAN} intelligence that targets near-real-time decision loops \cite{bonati2021intelligence}.

Despite progress on coverage, throughput, and cost objectives using optimization, heuristic, and learning-based approaches \cite{polese2020integrated,Madapatha2020IAB,simsek2021iab,shang2023reinforcement}, there remains limited consensus on how to (i) incorporate resilience objectives and uncertainty models in a scalable manner, (ii) enforce feasibility and constraints in learning-enabled planners, and (iii) evaluate robustness and transferability across scenarios using reproducible protocols. Robust optimization provides principled tools for uncertainty modeling, but can be conservative and computationally expensive at scale \cite{BenTal1998Robust,Bertsimas2011Robust}. Learning-based methods can be efficient at inference time, yet they may yield infeasible actions or generalize poorly without explicit structure and constraint handling \cite{mnih2015human,haarnoja2018soft,taylor2009transfer}. A central difficulty is that donor-topology variation does not merely change the geometric appearance of the deployment scenario; it also changes the initial coverage footprint, the location of transport bottlenecks, and the set of candidates that remain feasible as multi-hop resources are consumed. In other words, topology variation alters both the reward landscape and the feasible action space. This makes the planning problem fundamentally different from unconstrained coverage expansion: an algorithm must not only identify high-gain candidate sites, but also preserve transport feasibility and connectivity redundancy as the topology evolves. From a learning perspective, this leads to sparse success signals and strong sensitivity to scenario structure, which motivates the use of graph-structured state representations together with staged training rather than treating each topology as an isolated deployment case. The main contributions of this paper are summarized as follows:

\begin{itemize}
\item We model \gls{IAB} deployment as a constrained Markov decision process that couples access coverage with multi-hop transport feasibility, including hop-by-hop rate-budget degradation and an $m$-parent connectivity requirement for fault tolerance.

\item We develop a GATv2-PPO actor-critic framework that represents evolving deployments as attributed directed graphs and learns sequential node deployment decisions that capture donor-induced multi-hop dependencies under coverage and transport feasibility requirements.

\item We design a multi-stage curriculum that progressively increases constraint difficulty and scenario diversity, and we demonstrate transfer across heterogeneous donor topologies together with progressive backhaul-failure stress tests.
\end{itemize}

The remainder of this paper is organized as follows. Section~\ref{sec:litreview_tnse} reviews related work. Section~\ref{sec:sysForm} presents the system model and problem formulation. Section~\ref{sec:results} describes the simulation setup and discusses the results. Section~\ref{sec:conclusion} concludes the paper.

\section{Related Work}
\label{sec:litreview_tnse}

Research on \gls{IAB} network planning has evolved along two practical dimensions: scalability (from small instances to ultra-dense deployments) and dynamics (from quasi-static assumptions to time-varying \gls{mmWave} links). These requirements have shaped three main methodological directions. Exact optimization offers explicit feasibility guarantees but is difficult to scale. Heuristics and metaheuristics improve tractability but often rely on simplified surrogates and design rules. Learning-based approaches enable fast inference and adaptation, yet they introduce new challenges in constraint satisfaction and robustness. Across all three lines of work, resilience is increasingly recognized as essential: blockage and link degradation can disconnect multi-hop subtrees, so failure tolerance must be addressed during planning rather than treated solely as an operational concern \cite{Sterbenz2010Resilience,MacCartney2017}. Table~\ref{tab:related_work_summary} summarizes representative studies and positions this paper relative to the state of the art.

\subsection{Optimization-based planning and scalability limits}
Early studies on \gls{IAB} planning commonly used mathematical optimization to jointly model placement, association, and transport/resource constraints. These formulations couple deployment choices with end-to-end objectives and in-band resource multiplexing, providing interpretable constraints and strong benchmarks for moderate-scale settings \cite{polese2020integrated,Madapatha2020IAB}. However, computational complexity increases rapidly with the number of candidate sites, hop depth, and the coupling induced by interference and scheduling constraints, which limits direct applicability in ultra-dense scenarios. This observation is consistent with hardness results in related wireless mesh problems, such as joint routing and channel assignment \cite{Alicherry2006MobiCom}, as well as classical throughput scaling laws indicating that multi-hop wireless capacity can degrade quickly as hop count grows, making transport a first-order constraint in dense deployments \cite{Gupta2000Capacity}. Robust optimization provides principled tools to incorporate uncertainty, but robust mixed-integer formulations can be conservative and remain challenging to solve at large scale \cite{BenTal1998Robust,Bertsimas2011Robust}. Consequently, optimization-based approaches are most valuable for establishing reference formulations, deriving constraints, and benchmarking feasibility, while scalable planning typically requires additional approximations.
\subsection{Heuristics and metaheuristics as scalable approximations}

To handle large candidate sets, many studies employ
heuristics and evolutionary search to explore complex
feasibility regions with manageable runtime. Blocking-aware
constrained deployment optimization has been investigated
for IAB networks with geographical or interference-management
restrictions on node placement, demonstrating coverage gains
from network planning\cite{madapatha2022constrained}.

Our previous IAB planning study \cite{info15010019}
formulated single-hop deployment as a MILP and developed
a greedy approach for geographically constrained
multi-hop deployment. The single-hop Greedy--MILP
comparison provided an optimization reference for
evaluating the greedy approach. However, that study
focused on nominal coverage and data-rate feasibility;
it did not investigate learned cross-topology transfer
or explicitly evaluate deployment resilience under
backhaul-link failures. Multi-objective evolutionary methods, such as \gls{MOEA/D}, also provide a means of balancing competing
objectives when exact optimization is impractical
\cite{zhang2007moead}.
Despite their scalability, resilience is often handled
indirectly, for example through simple redundancy rules,
or evaluated post hoc. This weakens comparability across
studies because disruption models and resilience metrics
vary, and it risks producing deployments that perform
well under nominal assumptions but become vulnerable
under \gls{mmWave} link disruptions. 

\subsection{Learning-based planning and graph-structured representations}
Learning-based approaches seek to reduce online computation and improve adaptivity under non-stationary conditions. \gls{RL} is a natural fit because planning actions modify the network state and influence long-term performance \cite{sutton2018reinforcement}. In \gls{IAB} settings, learning has been applied to topology and association design, including approaches that combine graph embeddings with deep \gls{RL} to capture multi-hop interactions \cite{simsek2021iab}. Complementary work investigates decentralized or semi-centralized resource and path control, as well as scheduling, in multi-hop \gls{IAB}, with an emphasis on operational optimization under in-band constraints \cite{ALGHAFARI2022108837,singh2023distributed,lim2021joint}. A recurring issue is representation: non-graph approaches often require hand-crafted features to encode topology and hop dependencies. \glspl{GNN} mitigate this limitation by operating directly on graphs with permutation-invariant message passing \cite{scarselli2008graph,velivckovic2017graph}, and surveys in network management highlight their value for modeling and control tasks in which topology is fundamental \cite{liu2022graph}. Nonetheless, learning-based algorithms must still address hard feasibility constraints and robustness under distribution shifts.

\subsection{Resilience-aware gaps in \gls{IAB} planning}
Recent work on wireless backhaul management argues that resilience should be treated as a first-order objective through explicit failure models and reconfiguration mechanisms, rather than as a stress test applied after planning \cite{Gemmi2024ResilientIAB,Sterbenz2010Resilience}. In contrast, much of the \gls{IAB} planning literature continues to optimize primarily for coverage, throughput, and cost, while resilience is omitted or evaluated under simplified disruption assumptions.

\begin{table*}[!t]
\centering
\caption{Representative literature on \gls{IAB} planning/management, with emphasis on resilience treatment.}
\label{tab:related_work_summary}
\scriptsize
\renewcommand{\arraystretch}{1.18}
\setlength{\tabcolsep}{4pt}
\begin{tabular}{p{2.55cm} p{2.15cm} p{4.55cm} p{3.7cm} p{1.55cm}}
\toprule
\textbf{Reference} & \textbf{Method} & \textbf{Focus / contribution} & \textbf{What is missing for resilience-aware planning?} & \textbf{Resilience} \\
\midrule

\cite{polese2020integrated} &
Overview &
Architecture + design space; practical multi-hop mmWave challenges. &
No deployment optimizer; no explicit failure-aware constraints/objectives. &
None \\

\cite{Madapatha2020IAB} &
MIP/Optimization &
Joint deployment + access--backhaul coupling under end-to-end objectives. &
Scales poorly with dense candidates/hops; resilience not co-optimized. &
None \\

\cite{BenTal1998Robust,Bertsimas2011Robust} &
Robust theory &
Worst-case uncertainty modeling and feasibility guarantees. &
Conservative; robust MIP expensive at dense-network scale. &
Theory \\

\cite{madapatha2022constrained} &
Constrained planning &
Blocking-aware placement under geographical or interference-management restrictions. &
Coverage-focused comparison; not a learned cross-topology policy. &
Blockage-aware \\

\cite{info15010019} &
Greedy heuristic &
Geographically constrained single-hop and
multi-hop IAB planning; single-hop Greedy comparison
against a MILP benchmark. &
Resilience not a first-class objective; simplified disruption modeling. &
None \\

\cite{simsek2021iab} &
Graph+\gls{DRL} &
Learned representations for multi-hop topology/association. &
Hard constraints + transfer robustness not systematically ensured. &
None \\

\cite{ALGHAFARI2022108837, singh2023distributed, lim2021joint} &
Runtime management &
Path/resource control under in-band/half-duplex constraints. &
Assumes a given deployment; limited proactive resilience co-design. &
Runtime \\

\cite{Sterbenz2010Resilience} &
Survey &
Resilience concepts and metrics for communication networks. &
Not an IAB planning method; no deployment formulation/algorithm. &
Foundational \\

\cite{Gemmi2024ResilientIAB} &
Failure-aware control &
Motivates failure-aware control/reconfiguration for backhaul. &
Limited proactive deployment optimization under failures. &
Runtime \\

\midrule
\textbf{This paper} &
\textbf{GATv2 + curriculum \gls{RL}} &
\textbf{Constrained MDP for deployment: hop-by-hop transport budgeting + resilience;} &
\textbf{Requires offline training; depends on scenario/model fidelity.} &
\textbf{Planning} \\
\bottomrule
\end{tabular}
\end{table*}

\section{System Model and Problem Formulation}\label{sec:sysForm}

\subsection{System Model}
We consider an urban \gls{mmWave} \gls{IAB} deployment operating at 60~GHz. 
The service area is discretized into a grid of candidate installation sites representing existing street infrastructure, such as lamp posts and utility poles. Pole-based deployment has been considered in mmWave backhaul planning and in experimental studies of IAB radio channels~\cite{Hu2018RoadsideBackhaul,Askar2022LamppostIAB}.
Reusing street infrastructure for wireless nodes is also consistent with the deployment concepts discussed in the
NTT DOCOMO white paper on 5G evolution and
6G~\cite{NTTDOCOMO2021WhitePaper}.
 The network consists of fiber-connected \gls{IAB} donors $\mathcal{I}$ that provide gateways to the core network, and candidate \gls{IAB} node locations $\mathcal{J}$ that can be deployed to extend access coverage via wireless multi-hop backhaul.

A key modeling principle is that multi-hop expansion is governed by an end-to-end data-rate budget originating at the \gls{IAB} donors and degrading hop by hop. As traffic is forwarded from a donor to downstream hops, each intermediate \gls{IAB} node (i) consumes part of its incoming budget to serve local access demand and (ii) incurs protocol overhead captured by an overhead factor $R_o>1$. Consequently, the residual budget available for further forwarding decreases along any donor-rooted branch. This induces an upstream-to-downstream directionality and provides an implicit stopping condition: if the remaining budget at a hop cannot support the local access requirement after accounting for $R_o$, then further expansion along that branch is infeasible.

The above quantities are treated as design-stage inputs for
initial IAB deployment. This follows the common practice of
dimensioning wireless infrastructure using expected traffic
and capacity requirements rather than instantaneous traffic
realizations \cite{ciscoMerakiHighDensity}. The resulting
deployment is subsequently evaluated under time-varying
traffic conditions in Section~\ref{subsec:dynamic_traffic}.

\smallskip
\noindent\textbf{Link budget}
For any potential wireless link (donor--node or node--node), the received power (in dB units) is modeled as
\begin{equation}
P_r = P_{tx} + G_{\text{total}} - L_{\text{total}},
\label{eq:received_power}
\end{equation}
where $P_{tx}$ is the transmit power, $G_{\text{total}}$ is the total antenna gain, and $L_{\text{total}}$ is the aggregate propagation loss.

For access links serving user equipment, omni-directional antennas with constant gain $G_0$ are assumed:
\begin{equation}
G_{\text{access}} = G_0.
\label{eq:access_gain}
\end{equation}

For backhaul links, directional sector antennas are used. We adopt a simplified sectored-antenna model:
\begin{equation}
G_{\text{total}} =
\begin{cases}
G_t + G_r, & \text{if } |\theta| \leq \dfrac{\theta_{\text{HPBW}}}{2},\\[4pt]
G_{\text{side}}, & \text{otherwise},
\end{cases}
\label{eq:directional_gain}
\end{equation}
where $G_t$ and $G_r$ denote transmit and receive main-lobe gains, $\theta$ is the angular misalignment, $\theta_{\text{HPBW}}$ is the half-power beamwidth, and $G_{\text{side}}$ is a constant side-lobe gain.

The aggregate propagation loss at 60~GHz is
\begin{equation}
L_{\text{total}}(d) = L_{path}(d) + L_{atm}(d) + L_{rain}(d),
\label{eq:total_loss}
\end{equation}
where $d$ is the link distance, $L_{path}(d)$ denotes path loss, $L_{atm}(d)=O_{tr}\,d$ captures oxygen absorption, and $L_{rain}(d)=L_r\,d$ captures rain attenuation (with $O_{tr}$ and $L_r$ in dB/km and $d$ in km).

Link feasibility is determined by an \gls{SNR} threshold:
\begin{equation}
\text{SNR}_{pq} = P_r - N_0 \geq \text{SNR}_{\text{th}},
\label{eq:snr_feasibility}
\end{equation}
where the receiver noise power is
$N_0 = -174 + 10\log_{10}(W)$~dBm for bandwidth $W$ (Hz). For a feasible link, we map $\text{SNR}_{pq}$ to an achievable rate upper bound using Shannon capacity:
\begin{equation}
C_{pq} = \frac{W}{10^6}\log_2\!\left(1 + 10^{\text{SNR}_{pq}/10}\right)\ \ \text{(Mbps)},
\label{eq:link_rate_shannon}
\end{equation}
and set $C_{pq}=0$ if \eqref{eq:snr_feasibility} is not satisfied. Here, $C_{pq}$ represents the maximum usable backhaul data rate on link $p\!\rightarrow\!q$ under the adopted physical-layer assumptions.

\begin{table}[!t]
\centering
\caption{Representative parameter values for 60~GHz attenuation components.}
\label{tab:params}
\begin{tabular}{l c l}
\toprule
\textbf{Parameter} & \textbf{Value} & \textbf{Reference} \\
\midrule
$L_r$ (rain attenuation) & 12.6 dB/km & \cite{itu-r838} \\
$O_{tr}$ (oxygen absorption) & 16 dB/km & \cite{siles2015atmospheric} \\
$P_{r,\min}$ (coverage threshold) & $-80$ dBm & \cite{bai2015coverage} \\
\bottomrule
\end{tabular}
\end{table}

\subsection{Problem Formulation}
We formulate resilience-aware \gls{IAB} deployment as a mixed-integer optimization problem that minimizes the number of deployed \gls{IAB} nodes while enforcing coverage, connectivity, and resilience requirements. The coupling between topology and multi-hop transport feasibility is captured via hop-by-hop budget constraints that reflect protocol overhead and per-node access consumption.

\subsubsection{Decision Variables}
Let $\mathcal{I}$ denote the set of fiber-connected \gls{IAB} donors and $\mathcal{J}$ the set of candidate \gls{IAB} node locations. The service area is partitioned into grid cells indexed by $\mathcal{K}$.
The decision variables are:
\begin{itemize}
\item $\alpha_j \in \{0,1\}$: deployment indicator for candidate \gls{IAB} node $j \in \mathcal{J}$;
\item $Y_{pq} \in \{0,1\}$: backhaul link activation from node $p$ to node $q$, where $p \in \mathcal{I}\cup\mathcal{J}$ and $q \in \mathcal{J}$;
\item $R_{pq} \geq 0$: allocated payload rate (budget) on backhaul link $p \rightarrow q$ (Mbps);
\item $U_k \in \{0,1\}$: coverage indicator for grid cell $k \in \mathcal{K}$.
\end{itemize}
For all donors $i \in \mathcal{I}$, we set $\alpha_i = 1$.

\subsubsection{Coverage Constraints}
We precompute a binary coverage mask $\delta_{vk}\in\{0,1\}$ indicating whether transmitter $v\in \mathcal{I}\cup\mathcal{J}$ can cover cell $k\in\mathcal{K}$ under the adopted coverage criterion ($P_r\ge P_{r,\min}$). The coverage constraint is
\begin{equation}
U_k \leq \sum_{i\in\mathcal{I}} \delta_{ik} + \sum_{j\in\mathcal{J}} \delta_{jk}\alpha_j, \quad \forall k\in\mathcal{K}.
\label{eq:cell_coverage}
\end{equation}
The aggregate coverage requirement is
\begin{equation}
\sum_{k\in\mathcal{K}} U_k \geq \theta_{cov} |\mathcal{K}|,
\label{eq:total_coverage}
\end{equation}
ensuring that at least a fraction $\theta_{cov}$ of grid cells are covered, where $\theta_{\mathrm{cov}}$ specifies the minimum acceptable coverage fraction during planning.
\subsubsection{Resilience and Connectivity Constraints}

Backhaul activation obeys feasibility and deployment consistency:
\begin{align}
Y_{pq} &\leq L_{pq}, \nonumber\\
Y_{pq} &\leq \alpha_p, \nonumber\\
Y_{pq} &\leq \alpha_q,
\label{eq:link_activation}
\end{align}
for all $p\in\mathcal I\cup\mathcal J$ and
$q\in\mathcal J$ with $p\neq q$,
where $L_{pq}\in\{0,1\}$ indicates whether link $(p,q)$
is feasible according to \eqref{eq:snr_feasibility}.

Here, $Y_{pq}=1$ identifies a retained backhaul link
from an upstream IAB node or IAB donor $p$ to a
downstream IAB node $q$. A retained link need not
carry a positive allocated rate.

To provide backhaul redundancy, each deployed IAB node
must retain at least $m$ inbound backhaul links from
distinct upstream IAB nodes or IAB donors:

\begin{equation}
\sum_{p \in \mathcal{I}\cup\mathcal{J},\ p \neq j}
Y_{pj}
\geq m\,\alpha_j,
\quad \forall j\in\mathcal{J}.
\label{eq:resilience_constraint}
\end{equation}

The anti-parallel constraint excludes directly opposed
backhaul links between two IAB nodes:

\begin{equation}
Y_{pq}+Y_{qp}\leq 1,
\quad \forall p,q\in\mathcal J,\ p\neq q.
\label{eq:anti_parallel}
\end{equation}

To state the acyclic-topology requirement explicitly,
let $\mathfrak C_{\geq3}$ denote the family of simple
directed cycles containing at least three links in
the candidate backhaul graph defined by $L_{pq}=1$.
Each $\Gamma\in\mathfrak C_{\geq3}$ is represented
by its directed-link set, and $|\Gamma|$ is its
number of links. We impose
\begin{equation}
\sum_{(p,q)\in\Gamma}Y_{pq}
\leq |\Gamma|-1,
\quad \forall\Gamma\in\mathfrak C_{\geq3}.
\label{eq:cycle_elimination}
\end{equation}
This constraint prevents all links of any such cycle
from being retained simultaneously.

Under the upstream-to-downstream orientation adopted
for $Y_{pq}$, backhaul links terminate at IAB nodes
in $\mathcal J$, not at IAB donors. Every possible
directed cycle is therefore contained in $\mathcal J$.
Together, \eqref{eq:anti_parallel} and
\eqref{eq:cycle_elimination} ensure that the planned
backhaul topology is a directed acyclic graph (DAG).
The sequential deployment procedure satisfies these
conditions by construction, as explained in the
MDP transition description. The candidate backhaul
graph may contain cycles; the DAG requirement applies
to the retained links for which $Y_{pq}=1$.

\subsubsection{Capacity and Flow Constraints}

\noindent\textbf{(i) Link capacity with reservation.}
The allocated payload rate on an active link is
upper-bounded by the usable capacity while reserving
a fraction $\beta\in[0,1)$ for recovery under failures:
\begin{equation}
R_{pq}
\leq
(1-\beta)\,C_{pq}\,Y_{pq},
\quad
\forall p\in\mathcal I\cup\mathcal J,\ 
q\in\mathcal J,\ p\neq q.
\label{eq:link_capacity}
\end{equation}

\noindent\textbf{(ii) IAB donor budget constraint.}

For each IAB donor $i\in\mathcal I$, the outgoing
payload-rate budget and local access demand must
fit within its available fiber capacity after
accounting for overhead:

\begin{equation}
R_o\left(A_i+\sum_{j\in\mathcal J}R_{ij}\right)
\leq F_i,
\quad \forall i\in\mathcal I,
\label{eq:donor_capacity}
\end{equation}
where $F_i$ is the available fiber rate
at IAB donor $i$ and $A_i$ is its local access
demand, both expressed in Mbps.

\noindent\textbf{(iii) Hop-by-hop budget degradation.}
For each deployed IAB node $j\in\mathcal J$,
the total incoming payload budget must cover
its local access demand and the payload it
forwards downstream, both inflated by the
overhead factor $R_o$:
\begin{equation}
\sum_{\substack{p\in\mathcal I\cup\mathcal J\\p\neq j}}
R_{pj}
\geq
R_o\left(
A_j\alpha_j+
\sum_{\substack{n\in\mathcal J\\n\neq j}}R_{jn}
\right),
\quad \forall j\in\mathcal J.
\label{eq:flow_conservation}
\end{equation}

This constraint limits the residual forwarding budget
after accounting for local access demand and protocol
overhead. If it cannot be satisfied after adding a
candidate IAB node, that deployment extension is
infeasible under the rate-budget model.

For a single-parent forwarding configuration, let $p$
be the only upstream IAB node or IAB donor supplying
a positive allocated rate to deployed IAB node $j$.
Then $\sum_{u\neq j}R_{uj}=R_{pj}$.
For any outgoing link $j\rightarrow q$ with
$R_{jq}>0$, \eqref{eq:flow_conservation} gives
\begin{equation}
\begin{aligned}
R_{pj}&\geq R_o\left(A_j+\sum_{\substack{n\in\mathcal J\\n\neq j}}R_{jn}
\right)\\&\geq R_oR_{jq}>R_{jq},
\end{aligned}
\label{eq:single_parent_rate_decrease}
\end{equation}
because $\alpha_j=1$, $R_o>1$, and $A_j\geq0$.
A positive-rate directed cycle would therefore
require the allocated rate to decrease strictly
and return to its initial value, which is impossible.

This rate-based argument is conditional on
single-parent forwarding and concerns allocated
rates $R_{pq}$, not physical-link capacities
$C_{pq}$. Acyclicity of the complete planned
topology, including retained links with zero
allocated rate, is established separately by
\eqref{eq:anti_parallel},
\eqref{eq:cycle_elimination}, and the sequential
deployment construction.

\subsubsection{Optimization Objective}

The objective minimizes the number of deployed IAB nodes:
\begin{equation}
\min \sum_{j\in\mathcal J}\alpha_j
\label{eq:objective}
\end{equation}

subject to
\eqref{eq:cell_coverage}--\eqref{eq:flow_conservation},
including the anti-parallel and directed-cycle
elimination constraints. The formulation enforces
the coverage target, the minimum number of upstream
backhaul connections, link-capacity limits, and
hop-by-hop rate-budget feasibility within an
acyclic planned backhaul topology.

\subsection{Graph Representation}
\label{sec:graphForm}
To enable \gls{GNN}-based optimization, we map the above formulation to an attributed directed graph $\mathcal{G} = (\mathcal{V}, \mathcal{E}, \mathbf{X}, \mathbf{Z}, \mathbf{g})$ that encodes topology and deployment state.

\paragraph{Vertices}
$\mathcal{V} = \mathcal{I} \cup \mathcal{J}$ includes \gls{IAB} donors ($i \in \mathcal{I}$) and candidate \gls{IAB} nodes ($j \in \mathcal{J}$).

\paragraph{Directed edges}
$\mathcal{E} = \{(p, q)\mid L_{pq}=1,\ p \in \mathcal{I}\cup\mathcal{J},\ q \in \mathcal{J},\ p\neq q\}$ includes all feasible \gls{mmWave} backhaul links.

\paragraph{Node features}
$\mathbf{X}\in\mathbb{R}^{|\mathcal{V}|\times d_v}$ stacks node attributes
\[
\mathbf{x}_v=\Bigl[\alpha_v,\ \frac{A_v}{A_{\max}},\ \frac{N_v}{m},\ \mathbbm{1}_{\{v\in\mathcal{I}\}}\Bigr],
\]
where $\alpha_v$ is the deployment status, $A_v$ is the access demand, $A_{\max}=\max_{v\in\mathcal{V}}A_v$ is a normalization constant, $N_v=\sum_{p:(p,v)\in\mathcal{E}}Y_{pv}$ is the current number of active inbound parents, and $\mathbbm{1}_{\{v\in\mathcal{I}\}}$ indicates whether $v$ is a donor.

\paragraph{Edge features}
$\mathbf{Z}\in\mathbb{R}^{|\mathcal{E}|\times d_e}$ stacks directed-link attributes
\[
\mathbf{z}_{pq,t}=
\Bigl[\frac{C_{pq}}{C_{\max}},\ \frac{R_{pq,t}}{C_{pq}},\ L_{pq}\Bigr],
\]
where $C_{\max}=\max_{(p,q)\in\mathcal{E}}C_{pq}$, and $R_{pq,t}$ denotes the allocated rate on link $p\rightarrow q$ at decision step $t$.

\paragraph{Global features}
We use
\[
\mathbf{g}=\bigl[\theta_{cov},\ m,\ \beta,\ R_o\bigr]
\]
to encode the coverage target, resilience requirement, reserved headroom, and protocol overhead. This representation enables the \gls{GNN} to learn multi-hop dependencies while reasoning about deployment constraints and capacity utilization. This graph formulation is motivated by the observation that donor placements and evolving backhaul relations induce structured dependencies that are difficult to capture with flat feature vectors. In particular, whether a candidate site is beneficial depends not only on its local coverage gain, but also on the upstream transport state and the availability of feasible parent connections in the current deployment graph. Encoding the state as an attributed directed graph therefore allows the policy to condition its decisions on relational information that changes with topology, rather than relying solely on manually aggregated statistics.

\subsection{Markov Decision Process Formulation}

We formulate resilient \gls{IAB} deployment as an \gls{MDP} $(\mathcal{S},\mathcal{A},\mathcal{P},\mathcal{R},\gamma)$.

\paragraph{State Space $\mathcal{S}$}
The state at decision step $t$ is represented by the attributed graph $s_t := \mathcal{G}_t = (\mathcal{V}, \mathcal{E}, \mathbf{X}_t, \mathbf{Z}_t, \mathbf{g})$, where $\mathbf{X}_t$ and $\mathbf{Z}_t$ capture dynamic deployment and rate states, while $(\mathcal{V},\mathcal{E},\mathbf{g})$ remain fixed within a scenario.

\paragraph{Action Space $\mathcal{A}$}
The environment defines a discrete action space of size $N+1$. Actions $a_t\in\{0,\ldots,N-1\}$ deploy an \gls{IAB} node at the selected location, and a dedicated stop action $a_t=N$ terminates the episode. Importantly, the effective action set is state-dependent:
\[
\mathcal{A}(s_t) \;=\; \mathcal{J}_{\mathrm{feas}}(s_t)\ \cup\ \{\text{stop}\},
\]
where $\mathcal{J}_{\mathrm{feas}}(s_t)$ contains sites that are not yet deployed and admit at least $m$ feasible inbound parents. A parent is feasible if it is already deployed and has sufficient remaining data-rate (i.e., above an implementation threshold). The policy is trained with action masking over $\mathcal{A}(s_t)$ to handle the large discrete search space. In the present formulation, the policy is responsible for selecting the next node to deploy, whereas the assignment of inbound parents for a newly deployed node is handled by a deterministic environment-side feasibility routine. This design choice is intentional: jointly learning both node selection and parent-link assignment would lead to a substantially larger combinatorial action space, making stable policy optimization considerably more difficult under sparse feasibility signals. By separating deployment selection from parent assignment, the framework focuses learning on sequential topology expansion under constrained transport evolution, while still enforcing connectivity and budget consistency through the environment transition.

\paragraph{Transition Dynamics $\mathcal{P}$ and Implicit Rate-Limited Termination}
Let the environment state at step $t$ be the attributed directed graph
\(
\mathcal{G}_t=(\mathcal{V},\mathcal{E},\mathbf{X}_t,\mathbf{Z}_t,\mathbf{g})
\),
and let $a_t=j \in \mathcal{J}_{\mathrm{feas}}(s_t)$ denote the selected deployment action. The transition $\mathcal{P}$ is realized by augmenting the graph with a newly deployed vertex and its inbound backhaul edges, followed by a deterministic update of edge-rate and coverage attributes.

\smallskip
Deploying node $j$ updates the vertex deployment indicator:
\begin{equation}
\alpha_{j,t+1}=1,\qquad \alpha_{v,t+1}=\alpha_{v,t}\ \ \forall v\neq j .
\label{eq:alpha_update_v2}
\end{equation}
Let $\mathcal{V}^{\mathrm{dep}}_t=\{v\in\mathcal{V}\mid \alpha_{v,t}=1\}$ denote the deployed set at time $t$ and let $\tilde{F}_{p,t}$ be the remaining forwarding budget at an upstream node $p\in\mathcal{V}^{\mathrm{dep}}_t$. The environment selects an inbound parent set $\mathcal{P}_{j,t}\subseteq \mathcal{V}^{\mathrm{dep}}_t\!\setminus\!\{j\}$ with $|\mathcal{P}_{j,t}|=m$ using Algorithm~\ref{alg:parent_selection}. The corresponding directed edges are activated:
\begin{equation}
\begin{aligned}
Y_{p j,t+1}&=1,\qquad \forall p\in\mathcal{P}_{j,t},\\
Y_{pq,t+1}&=Y_{pq,t},\\
&\quad \forall (p,q)\notin\{(p,j)\mid p\in\mathcal{P}_{j,t}\}.
\end{aligned}
\label{eq:edge_activation_v2}
\end{equation}

\smallskip
\noindent\textbf{(ii) Attribute update under hop-by-hop budgeting.}
Given the updated deployment and edge set, the edge-rate attributes are updated by a deterministic budgeting operator $\Phi(\cdot)$ consistent with the hop-by-hop degradation model:
\begin{equation}
\mathbf{Z}_{t+1} \;=\; \Phi\!\left(\mathcal{G}_t,\ j,\ \mathcal{P}_{j,t}\right),
\label{eq:z_update_v2}
\end{equation}
and the node attributes (including coverage-related features) are recomputed accordingly:
\begin{equation}
\mathbf{X}_{t+1} \;=\; \Psi\!\left(\mathcal{G}_t,\ j,\ \mathcal{P}_{j,t},\ \mathbf{Z}_{t+1}\right).
\label{eq:x_update_v2}
\end{equation}
Therefore, the next state is
\begin{equation}
\mathcal{G}_{t+1} \;=\; (\mathcal{V},\mathcal{E},\mathbf{X}_{t+1},\mathbf{Z}_{t+1},\mathbf{g}),
\label{eq:state_transition_v2}
\end{equation}
where $(\mathcal{V},\mathcal{E},\mathbf{g})$ are scenario-fixed, and $(\mathbf{X}_t,\mathbf{Z}_t)$ evolve with the deployment.

The sequential transition preserves acyclicity.
Initially, the deployed topology contains only
IAB donors and no planned backhaul links.
At decision step $t$, the selected IAB node $j$
is not yet deployed, and its upstream set satisfies
$\mathcal P_{j,t}\subseteq
\mathcal V_t^{\mathrm{dep}}\setminus\{j\}$.
According to \eqref{eq:edge_activation_v2}, the
transition adds only links $p\rightarrow j$ from
already deployed IAB nodes or existing IAB donors;
all other planned links remain unchanged.

\smallskip
\emph{Implicit rate-limited termination:} As the deployment grows, per-node access consumption and protocol overhead reduce the remaining forwarding budgets $\tilde{F}_{p,t}$, which in turn shrinks the feasible action set $\mathcal{J}_{\mathrm{feas}}(s_t)$. Therefore, multi-hop expansion terminates naturally when no candidate admits $m$ valid parents with sufficient remaining budget, or when the agent selects the stop action after meeting the success criteria.

\begin{algorithm}[t]
\caption{Inbound Parent Selection for Deployed Node $j$}
\label{alg:parent_selection}
\begin{algorithmic}[1]
\REQUIRE Current graph state $\mathcal{G}_t$, selected node $j$, minimum in-degree $m$
\ENSURE Parent set $\mathcal{P}_{j,t}$

\STATE $\mathcal{V}^{\mathrm{dep}}_t \leftarrow \{v\in\mathcal{V}\mid \alpha_{v,t}=1\}$
\STATE $\mathcal{U}_t \leftarrow \mathcal{V}^{\mathrm{dep}}_t \setminus \{j\}$ \COMMENT{Upstream candidates}
\STATE $\mathcal{U}_t \leftarrow \{p\in\mathcal{U}_t \mid (p,j)\in\mathcal{E}\}$ \COMMENT{Feasible directed links}
\STATE For each $p\in\mathcal{U}_t$, compute remaining budget $\tilde{F}_{p,t}$ from $\mathbf{Z}_t$ (and/or cached node budgets)
\STATE Sort $\mathcal{U}_t$ in descending order of $\tilde{F}_{p,t}$
\STATE $\mathcal{P}_{j,t} \leftarrow \textsc{TopM}(\mathcal{U}_t, m)$
\RETURN $\mathcal{P}_{j,t}$
\end{algorithmic}
\end{algorithm}

\paragraph{Reward Function $\mathcal{R}$}
We adopt an objective-aligned reward that encourages rapid coverage improvement while minimizing the number of deployed \gls{IAB} nodes. Let $\mathcal{C}_t \in [0,1]$ denote the covered-cell ratio at step $t$, and define the incremental coverage gain as $\Delta \mathcal{C}_t \triangleq \mathcal{C}_{t+1}-\mathcal{C}_t$.

The per-step reward is
\begin{equation}
r_t
=
w_{\mathrm{cov}}\,\Delta \mathcal{C}_t
-
c_{\mathrm{dep}}\;\mathbbm{1}\{a_t \neq \text{stop}\}
+
r_t^{\mathrm{term}},
\label{eq:reward_s1}
\end{equation}
where $w_{\mathrm{cov}}>0$ weights coverage progress and $c_{\mathrm{dep}}>0$ imposes a constant cost per deployment action, promoting deployment efficiency.

Episodes terminate when the agent selects the stop action or when no feasible deployment actions remain. The terminal reward is
\begin{equation}
r_t^{\mathrm{term}} =
\begin{cases}
B_{\mathrm{s}} - \lambda\,n_T, & \text{if success},\\[2pt]
-\Bigl(B_{\mathrm{f}} + B_{\mathrm{g}}\,\mathrm{gap}\Bigr), & \text{otherwise},
\end{cases}
\label{eq:terminal_reward_s1}
\end{equation}
where $n_T$ is the total number of deployed \gls{IAB} nodes at termination, $\mathrm{gap}\triangleq [\theta_{\mathrm{cov}}-\mathcal{C}_T]_+$ is the coverage shortfall with $[x]_+\triangleq \max\{x,0\}$, and $B_{\mathrm{s}},B_{\mathrm{f}},B_{\mathrm{g}},\lambda>0$ are scalar coefficients. Success is declared when $\mathcal{C}_T \ge \theta_{\mathrm{cov}}$ and the final deployment satisfies all feasibility requirements enforced by the environment (e.g., minimum inbound connectivity and transport feasibility).

\begin{algorithm}[htbp]
\caption{Deployment rollout with a fixed GATv2--PPO policy}
\label{GATv2-PPO}
\begin{algorithmic}[1]
\REQUIRE IAB donor layout $\mathcal{D}$, candidate sites $\mathcal{J}$, stage constraints, fixed rollout actor--critic parameters $\theta_{\mathrm{old}}$
\ENSURE Episode trajectory $\tau$, return $R$, and final deployment
\STATE Reset environment; set $s_0$, $t\gets0$, $d\gets\mathrm{false}$, $\tau\gets\emptyset$
\WHILE{$d=\mathrm{false}$}
    \STATE Construct $\mathcal{G}_t$ and the feasible-action mask $\mathcal{A}(s_t)$
    \STATE Encode $\mathcal{G}_t$ using GATv2 and evaluate the masked policy $\pi_{\theta_{\mathrm{old}}}$
    \STATE Sample $a_t\sim\pi_{\theta_{\mathrm{old}}}(\cdot|s_t)$
    \STATE Execute $a_t$; obtain $s_{t+1}$, reward $r_t$, and terminal flag $d$ using Eqs.~\eqref{eq:reward_s1}--\eqref{eq:terminal_reward_s1}
    \STATE Store $(s_t,a_t,r_t,s_{t+1},d)$, mask, old log-probability, and old value estimate in $\tau$
    \STATE $t\gets t+1$
\ENDWHILE
\STATE $R\gets\sum_{t=0}^{|\tau|-1}r_t$
\RETURN $\tau$, $R$, and final deployment
\end{algorithmic}
\end{algorithm}
Algorithm~\ref{GATv2-PPO} collects one rollout; it does not update the policy. The stop action or exhaustion of feasible deployment actions terminates the episode. Algorithm~\ref{alg:curriculum_training} performs the training updates.
\paragraph{GATv2-based Policy Architecture}
The policy employs a GATv2 encoder with edge-conditioned attention to process the graph state:
\[
\mathbf{h}_v^{(l+1)} = \text{GATv2}^{(l)}(\mathbf{h}_v^{(l)}, \{\mathbf{h}_u^{(l)}, \mathbf{e}_{uv}\}_{u \in \mathcal{N}(v)})
\]
where $\mathbf{h}_v^{(l)}$ represents node embeddings at layer $l$, and edge features $\mathbf{e}_{uv}$ condition the attention mechanism. A pointer-based actor network computes deployment probabilities over valid candidate locations, while a critic network estimates state values for \gls{PPO} training.

\begin{table}[htbp]
\centering
\caption{Key Symbols and Definitions}
\label{tab:notation}
\resizebox{0.47\textwidth}{!}{
\begin{tabular}{c p{7.2cm}}
\hline
\textbf{Symbol} & \textbf{Description} \\
\hline

\multicolumn{2}{c}{\textit{Sets}} \\
\hline
$\mathcal{I}$ & Set of fiber-connected \gls{IAB} donor locations \\
$\mathcal{J}$ & Set of candidate locations for deployable \gls{IAB} nodes \\
$\mathcal{K}$ & Set of grid cells in the service area \\
$\mathcal{V}$ & Set of graph vertices, with $\mathcal{V}=\mathcal{I}\cup\mathcal{J}$ \\
$\mathcal{E}$ & Set of feasible directed backhaul edges in the graph representation \\
\hline

\multicolumn{2}{c}{\textit{Propagation and Link Parameters}} \\
\hline
$P_{\text{tx}}$ & Transmit power (dBm) \\
$P_r$ & Received power (dBm) \\
$G_0$ & Omni-directional antenna gain for access links (dB) \\
$G_t, G_r$ & Transmit and receive main-lobe antenna gains for backhaul links (dB) \\
$G_{\text{total}}$ & Total antenna gain used in the received-power calculation (dB) \\
$G_{\text{side}}$ & Side-lobe gain (dB) \\
$\theta$ & Angular misalignment between transmitter and receiver boresights (degrees) \\
$\theta_{\text{HPBW}}$ & Half-power beamwidth (degrees) \\
$L_{path}(d)$ & Distance-dependent path loss at link distance $d$ (dB) \\
$L_{atm}(d)$ & Atmospheric attenuation at distance $d$ (dB) \\
$L_{rain}(d)$ & Rain attenuation at distance $d$ (dB) \\
$L_{\text{total}}(d)$ & Total propagation loss at distance $d$ (dB) \\
$O_{tr}$ & Oxygen absorption coefficient (dB/km) \\
$L_r$ & Rain attenuation coefficient (dB/km) \\
$W$ & System bandwidth (Hz) \\
$N_0$ & Receiver noise power (dBm) \\
$\text{SNR}_{pq}$ & Signal-to-noise ratio of link $p \rightarrow q$ (dB) \\
$\text{SNR}_{\text{th}}$ & Minimum SNR threshold for backhaul feasibility \\
$C_{pq}$ & Achievable backhaul rate upper bound of link $p \rightarrow q$ (Mbps) \\
$L_{pq}$ & Binary indicator of whether directed link $(p,q)$ is feasible \\
\hline

\multicolumn{2}{c}{\textit{Coverage and Demand Parameters}} \\
\hline
$P_{r,\min}$ & Minimum received-power threshold for coverage (dBm) \\
$\theta_{\mathrm{cov}}$ & Minimum required coverage ratio \\
$\delta_{vk}$ & Binary coverage mask indicating whether transmitter $v \in \mathcal{I}\cup\mathcal{J}$ can cover grid cell $k \in \mathcal{K}$ \\
$U_k$ & Binary indicator of whether grid cell $k$ is covered \\
$A_i, A_j$ & Local access demand at donor $i$ or candidate/deployed node $j$ (Mbps) \\
$\mathcal{C}_t$ & Covered-cell ratio at decision step $t$ \\
\hline

\multicolumn{2}{c}{\textit{Decision Variables and Constraints}} \\
\hline
$\alpha_j$ & Binary deployment indicator for candidate node $j \in \mathcal{J}$ \\
$Y_{pq}$ & Binary indicator of whether the directed backhaul link from $p$ to $q$ is activated, where $p \in \mathcal{I}\cup\mathcal{J}$, $q \in \mathcal{J}$, and $p \neq q$ \\
$R_{pq}$ & Allocated payload rate on active backhaul link $p \rightarrow q$ (Mbps) \\
$m$ & Minimum required number of inbound parent connections for each deployed \gls{IAB} node \\
$\beta$ & Reserved backhaul-capacity fraction for robustness under failures \\
$F_i$ & Available fiber rate at donor $i$ (Mbps) \\
$R_o$ & Protocol-overhead factor for forwarding and local service \\
\hline

\multicolumn{2}{c}{\textit{Graph / MDP / Learning Variables}} \\
\hline
$\mathbf{X}$ & Node-feature matrix in the graph representation \\
$\mathbf{Z}$ & Edge-feature matrix in the graph representation \\
$\mathbf{g}$ & Global feature vector in the graph representation \\
$\mathcal{G}_t$ & Graph-structured state at decision step $t$ \\
$\mathcal{A}(s_t)$ & Feasible action set at state $s_t$ \\
$a_t$ & Action selected at decision step $t$ \\
$r_t$ & Reward at decision step $t$ \\
$\gamma$ & Discount factor in the Markov decision process \\
\hline
\end{tabular}}
\end{table}

\begin{figure*}[!t]
\centering
\includegraphics[width=\textwidth]{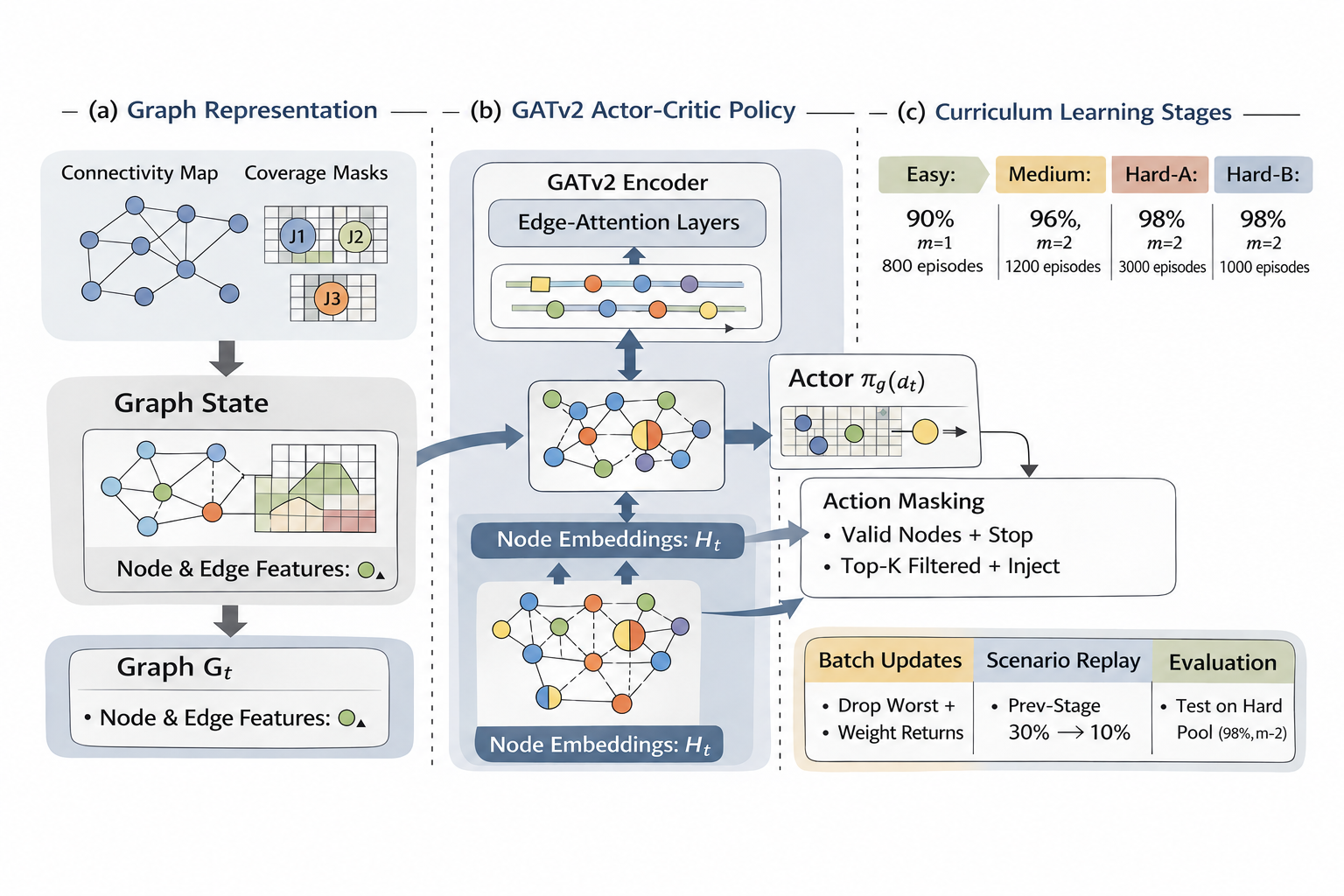}
\caption{Overview of the proposed GATv2--PPO framework
with curriculum learning for resilient IAB deployment.
(a) Graph-state construction from static connectivity and
coverage information together with dynamic deployment
attributes. (b) GATv2-based actor--critic with action masking
over the state-dependent feasible action set. (c) Four-stage
curriculum with progressively tighter planning requirements:
Easy ($\theta_{\mathrm{cov}}=90\%$, $m=1$),
Medium ($96\%$, $m=2$), Hard-A ($98\%$, $m=2$),
and Hard-B ($98\%$, $m=2$).}
\label{fig:network_deployment}
\end{figure*}

\subsection{Curriculum Learning Framework}
\label{sec:training_strategy}

Strict coverage, transport, and redundant-connectivity requirements make successful deployment episodes difficult to obtain early in training. The curriculum addresses this feasibility-learning problem by first exposing the policy to simpler IAB donor layouts and relaxed constraints, then tightening the requirements and diversifying the layouts. The intended benefit is to retain useful deployment decisions as the feasible action set becomes more restricted. The stage design is specified below; its deployment-level effect is examined against Direct-Hard training in Section~\ref{subsec:efficiency}.

\paragraph{Progressive Difficulty Scaling}
Our curriculum decomposes the deployment task into four
sequential stages, each defined by a coverage target
$\theta_{\text{cov}}$, a minimum parent-connectivity
requirement $m$, and the complexity of the IAB donor
scenario pool.
The corresponding stage targets are
$(90\%,1)$, $(96\%,2)$, $(98\%,2)$, and $(98\%,2)$
for Easy, Medium, Hard-A, and Hard-B, respectively,
consistent with Table~\ref{tab:curriculum_stages}
and Fig.~\ref{fig:network_deployment}. Table~\ref{tab:curriculum_stages} summarizes the stage specifications. The curriculum follows three principles of gradual complexity increase: (i) coverage targets increase from 90\% to 98\%, (ii) resilience constraints transition from single connectivity ($m=1$) to dual redundancy ($m=2$), and (iii) scenario pools expand from regular geometric \gls{IAB} donor patterns to irregular topologies.

\begin{table}[t]
\centering
\caption{Curriculum stage specifications and training parameters.}
\label{tab:curriculum_stages}
\begin{tabular}{lccccc}
\toprule
\textbf{Stage} & $\theta_{\text{cov}}$ & $m$ & \textbf{Episodes} & \textbf{LR} & $\lambda_{\text{ent}}$ \\
\midrule
Easy & 90\% & 1 & 800 & $3 \times 10^{-4}$ & $0.010 \to 0.004$ \\
Medium & 96\% & 2 & 1200 & $2 \times 10^{-4}$ & $0.006 \to 0.002$ \\
Hard-A & 98\% & 2 & 3000 & $1 \times 10^{-4}$ & $0.003 \to 0.0012$ \\
Hard-B & 98\% & 2 & 1000 & $5 \times 10^{-5}$ & $0.0018 \to 0.0010$ \\
\bottomrule
\end{tabular}
\end{table}

Stage~1 (Easy) establishes baseline deployment competence under relaxed constraints, using regular \gls{IAB} donor configurations (Pentagon, Hexagon, and Five-Dice) with single-connectivity resilience. Stage~2 (Medium) introduces full dual-connectivity requirements while moderately increasing the coverage target, and expands the scenario pool to include linear donor arrangements. Stage~3 (Hard-A) matches the final 98\% coverage specification and emphasizes consistent constraint satisfaction across diverse topologies, including irregular Sparse-Random scenarios. Stage~4 (Hard-B) maintains the target constraints and further refines the policy using a smaller learning rate and more conservative exploration.

\paragraph{Soft Curriculum Mixing}
Abrupt transitions between stages can lead to catastrophic forgetting. To mitigate this effect, we employ adaptive scenario replay in the Medium and Hard stages. At training episode $e$ within a stage of duration $E$, scenarios from the previous stage are sampled with probability
\begin{equation}
p_{\text{replay}}(e) = p_{\text{start}} + \frac{e-1}{E-1}\left(p_{\text{end}} - p_{\text{start}}\right),
\label{eq:replay_schedule}
\end{equation}
where $e=1,\ldots,E$ and $E>1$. For Stage~3, $p_{\text{start}} = 0.35$ and $p_{\text{end}} = 0.10$. This schedule maintains 30\% exposure to Medium-stage scenarios at the beginning of Stage~3 and linearly decays it to 10\% by the end of the stage, enabling a gradual transition while preserving previously learned skills.

\paragraph{Disaster Isolation Mechanism}
Policy-gradient updates can be destabilized by catastrophically poor episodes caused by early exploration or unfavorable scenario instances. We therefore adopt a multi-episode update protocol with worst-episode filtering to improve training stability. Experiences are accumulated over $B_{\text{ep}}$ episodes ($B_{\text{ep}} = 4$ for Easy/Medium and $B_{\text{ep}} = 8$ for the Hard stages) before each PPO update. Within each batch, episodes are ranked by cumulative return $R_i$, the single worst episode is discarded, and the remaining episodes are weighted by their normalized returns:

Let $\mathcal I_{\mathrm{keep}}$ denote the retained episodes,
$R_{\mathrm{lo}}=\min_{j\in\mathcal I_{\mathrm{keep}}}R_j$,
and $R_{\mathrm{hi}}=\max_{j\in\mathcal I_{\mathrm{keep}}}R_j$.
We use
\begin{equation}
w_i=
\begin{cases}
0.2+0.8\,\dfrac{R_i-R_{\mathrm{lo}}}{R_{\mathrm{hi}}-R_{\mathrm{lo}}},
& R_{\mathrm{hi}}>R_{\mathrm{lo}},\\[4pt]
1, & R_{\mathrm{hi}}=R_{\mathrm{lo}}.
\end{cases}
\label{eq:episode_weight}
\end{equation}
Thus $w_i\in[0.2,1]$; equal returns receive equal weights
without division by zero. Let $\mathcal T$ contain the
retained transitions and let $i(t)$ identify the episode
containing transition $t$. Using a minimization convention,
the actor loss is
\begin{equation}
\begin{aligned}
L_{\mathrm{actor}}
&=-\frac{1}{|\mathcal T|}
\sum_{t\in\mathcal T}w_{i(t)}
\min\!\bigl\{
\rho_t(\theta)\hat A_t,\\[-2pt]
&\hspace{35mm}
\operatorname{clip}(\rho_t(\theta),1-\epsilon,1+\epsilon)
\hat A_t
\bigr\},
\end{aligned}
\label{eq:weighted_loss}
\end{equation}
where
$\rho_t(\theta)=
\pi_\theta(a_t|s_t)/\pi_{\theta_{\mathrm{old}}}(a_t|s_t)$
is the probability ratio, distinct from reward $r_t$,
and $\hat A_t$ is the generalized advantage estimate.
Old action probabilities, realized feasible-action masks,
advantages, and value targets are held fixed during each
PPO update. If the retained episodes' mean return is below
the stage-specific update threshold, the update is skipped.

\paragraph{Exploration Scheduling}
Entropy regularization is used to balance exploration and exploitation. The entropy coefficient is decayed linearly within each stage using the schedule in Eq.~\eqref{eq:replay_schedule}, with stage-specific ranges reported in Table~\ref{tab:curriculum_stages}. The resulting PPO loss to be minimized is
\begin{equation}
L_{\text{total}} = L_{\text{actor}} - \lambda_{\text{ent}}(e)\,\mathcal{H}[\pi_\theta(\cdot|s)] + 0.5\,L_{\text{critic}},
\label{eq:ppo_total}
\end{equation}
where $\mathcal{H}[\pi_\theta]$ denotes the policy entropy. Earlier stages use higher entropy to encourage broad exploration, whereas later stages reduce entropy to promote exploitation of learned deployment patterns.

\paragraph{Unified Model Selection}
To limit stage-specific overfitting, we use a fixed evaluation protocol throughout training. Every 200 training episodes, the current policy is evaluated on all Hard-stage scenarios using the 98\% coverage target and $m=2$ resilience requirement, with three random seeds per scenario. Policies are ranked lexicographically by success rate (fraction of episodes satisfying all constraints), deployment efficiency (mean number of deployed \gls{IAB} nodes among successful episodes), and average achieved coverage. 

\begin{algorithm}[t]
\caption{Curriculum training with stability mechanisms.}
\label{alg:curriculum_training}
\begin{algorithmic}[1]
\REQUIRE Curriculum stages $\{\mathcal{S}_1, \ldots, \mathcal{S}_4\}$, agent $\pi_\theta$
\ENSURE Optimized policy $\pi_\theta^*$
\STATE {Initialize actor--critic parameters $\theta$; best validation score $\gets$ undefined}
\FOR{stage $k = 1$ to $4$}
    \STATE Set learning rate $\eta_k$ and entropy schedule from Table~\ref{tab:curriculum_stages}
    \STATE Episode buffer $\mathcal{B} \gets \emptyset$; {$\theta_{\mathrm{old}}\gets\theta$}
    \FOR{episode $e = 1$ to $E_k$}
        \STATE Sample a scenario (with replay probability from Eq.~\eqref{eq:replay_schedule} {when stage replay is enabled})
        \STATE {Run Algorithm~\ref{GATv2-PPO} with $\theta_{\mathrm{old}}$ to obtain the trajectory and $R_e$}
        \STATE $\mathcal{B} \gets \mathcal{B} \cup \{(\text{episode}, R_e)\}$
        \IF{$|\mathcal{B}| = B_{\text{ep}}$ \textbf{or} $e = E_k$}
            \STATE {Rank the $|\mathcal B|$ episodes by return; discard the worst if $|\mathcal B|>1$}
            \IF{{retained mean return meets the stage-specific update threshold}}
                \STATE Compute weights $\{w_i\}$ via Eq.~\eqref{eq:episode_weight}
                \STATE {Compute fixed advantages and value targets from the stored rollout data}
                \STATE {Minimize Eq.~\eqref{eq:ppo_total} using the actor loss in Eq.~\eqref{eq:weighted_loss} over the specified PPO epochs}
            \ENDIF
            \STATE $\mathcal{B} \gets \emptyset$; {$\theta_{\mathrm{old}}\gets\theta$}
        \ENDIF
        \IF{$e \bmod 200 = 0$}
            \STATE Evaluate on the fixed Hard-stage scenarios at the 98\% threshold
            \IF{{first validation or better lexicographic score: higher success, fewer IAB nodes, then higher coverage}}
                \STATE best score $\gets$ current score; save $\theta$ to memory
            \ENDIF
        \ENDIF
    \ENDFOR
\ENDFOR
\RETURN best $\pi_\theta^*$
\end{algorithmic}
\end{algorithm}

Algorithm~\ref{alg:curriculum_training} summarizes the complete training procedure, integrating progressive difficulty scaling, soft curriculum mixing, disaster isolation, and unified model selection. On a single NVIDIA RTX 4090 GPU, the full curriculum (6000 episodes) takes approximately 10 hours using mixed-precision training and cached graph structures.

\section{Main Results and Discussion}
\label{sec:results}

\subsection{Experimental Setup Overview}
\label{subsec:setup}
In our experimental evaluation, we simulate a 60\,GHz urban environment covering an area of \(1000\,\mathrm{m}\times1000\,\mathrm{m}\). The service area is discretized into a grid, with candidate \gls{IAB} node locations spaced at 50\,m intervals. The wireless propagation model adopts key parameters from \gls{3GPP} TR~38.901, including a 60\,GHz carrier frequency, high free-space path loss with additional atmospheric attenuation, and directional beamforming gains on the backhaul links. A grid cell is considered covered when the received power exceeds \(-80\)~dBm \cite{huawei2023linkbudget}. This threshold defines the nominal spatial coverage
criterion used for deployment planning; operational traffic
support is evaluated separately under time-varying demand. More specific coefficient values are reported in Table~\ref{tab:appendix_full}.

All experiments target a stringent coverage requirement of $\theta_{\mathrm{cov}} = 98\%$ and enforce a dual-connectivity resilience constraint ($m = 2$). To compare algorithms under different backhaul starting conditions, we evaluate multiple \gls{IAB} donor-placement scenarios. Each scenario specifies a fixed set of fiber-connected \gls{IAB} donor sites, as described below:

\textbf{Five-Dice}: one \gls{IAB} donor at the grid center and four near quarter-grid points;

\textbf{Pentagon}: five \gls{IAB} donors placed on a ring around the center with evenly spaced angles;

\textbf{Hexagon}: \gls{IAB} donors placed on a ring with evenly spaced angles, forming a hexagon-like geometry;

\textbf{Vertical} and \textbf{Horizontal}: \gls{IAB} donors aligned along a vertical/horizontal line through the center with symmetric offsets;

\textbf{Sparse-Random}: \gls{IAB} donors sampled randomly subject to a minimum-separation constraint;

\textbf{Dense-Cluster}: \gls{IAB} donors concentrated within a compact local region.

These scenarios are not cosmetic variations: they directly affect (i) the initial coverage footprint and (ii) the early-stage feasible action space. Under dispersed or near-symmetric donor layouts, the initial coverage is typically higher, leaving fewer uncovered grid cells; consequently, many candidate actions yield only marginal gains, and the algorithm must avoid redundant \gls{IAB} node deployments while terminating efficiently. Under clustered or strongly anisotropic layouts, the initial coverage is typically lower and uncovered regions are larger, which increases the number of actions that can improve coverage but also makes feasible expansion patterns more fragile as transport resources deplete across hops. Evaluating across diverse donor topologies therefore provides a diagnostic test of whether an algorithm remains effective when the feasible set and coverage gradients change substantially, which is central to the generalization objective of this work.
\begin{table*}[!t]
\centering
\caption{Complete Configuration Used in This Work.}
\label{tab:appendix_full}
\scriptsize
\setlength{\tabcolsep}{4pt}
\renewcommand{\arraystretch}{1.12}
\begin{tabularx}{\textwidth}{l X l l}
\hline
\textbf{Block} & \textbf{Item} & \textbf{Symbol / Name} & \textbf{Value} \\
\hline

\multicolumn{4}{c}{\textit{Reward Function} (Eqs.~\eqref{eq:reward_s1}--\eqref{eq:terminal_reward_s1})} \\
\hline
Reward & Coverage progress weight & $w_{\mathrm{cov}}$ & 10.0\textsuperscript{$\dagger$} \\
Reward & Deployment cost per non-stop action & $c_{\mathrm{dep}}$ & 0.1\textsuperscript{$\dagger$} \\
Reward & Success bonus (terminal) & $B_{\mathrm{s}}$ & 50.0\textsuperscript{$\dagger$} \\
Reward & Failure base penalty (terminal) & $B_{\mathrm{f}}$ & 20.0\textsuperscript{$\dagger$} \\
Reward & Gap penalty weight (terminal) & $B_{\mathrm{g}}$ & 100.0\textsuperscript{$\dagger$} \\
Reward & Node penalty at success (total deployed IAB nodes) & $\lambda$ & 1.0\textsuperscript{$\dagger$} \\

\hline

\multicolumn{4}{c}{\textit{Curriculum Learning: Stages, Targets, and Topology Pools} (Sec.~\ref{subsec:curriculum_hp_results})} \\
\hline
Metric & Primary training/evaluation metric & -- & success rate\textsuperscript{$\dagger$} \\
Target & Final strict coverage target & $\theta_{\mathrm{cov}}$ & 98\%\textsuperscript{$\dagger$} \\
Constraint & Minimum parent-connectivity requirement & $m$ & stage-dependent\textsuperscript{$\dagger$} \\
Curriculum & Number of stages & -- & 4 (Easy / Medium / Hard-A / Hard-B)\textsuperscript{$\dagger$} \\

Stage (Easy) &
Relaxed constraint profile to enlarge feasible set; regular donor geometries &
$(\theta_{\mathrm{cov}},m)$; pool &
(90\%,1); Five-Dice, Pentagon, Hexagon\textsuperscript{$\dagger$} \\

Stage (Medium) &
Enforces target feasibility requirements; introduces anisotropic layouts &
$(\theta_{\mathrm{cov}},m)$; pool &
(96\%,2); + Vertical, Horizontal\textsuperscript{$\dagger$} \\

Stage (Hard-A) &
Maintains final target; incorporates irregular donor placement &
$(\theta_{\mathrm{cov}},m)$; pool &
(98\%,2); + Sparse-Random\textsuperscript{$\dagger$} \\

Stage (Hard-B) &
Final refinement under the same hard constraint profile &
$(\theta_{\mathrm{cov}},m)$; pool &
(98\%,2); + Sparse-Random\textsuperscript{$\dagger$} \\
\hline

\multicolumn{4}{c}{\textit{Model and PPO Optimization }} \\
\hline
GNN & GATv2 layers / hidden / heads / dropout & $(L,d_h,n_{\mathrm{heads}},p_{\mathrm{drop}})$ &
(3, 128, 4, 0.1)\textsuperscript{$\dagger$} \\
GNN & Output embedding dimension & $d_{\mathrm{out}}$ & 64\textsuperscript{$\dagger$} \\
PPO & Discount factor / GAE parameter & $(\gamma,\lambda_{\mathrm{GAE}})$ &
(0.99, 0.95)\textsuperscript{$\dagger$} \\
PPO & Clipping range / value loss coefficient & $(\epsilon,c_v)$ &
(0.2, 0.5)\textsuperscript{$\dagger$} \\
PPO & Max gradient norm & $g_{\max}$ & 0.5\textsuperscript{$\dagger$} \\
PPO & Mini-batch size / PPO epochs & $(B_{\mathrm{mini}},K_{\mathrm{epoch}})$ &
(64, 4)\textsuperscript{$\dagger$} \\
PPO & Optimizer & -- & Adam\textsuperscript{$\dagger$} \\
\hline

\end{tabularx}

\vspace{2pt}
\parbox{\textwidth}{\scriptsize
\textsuperscript{$\dagger$}Values and Eqs.~\eqref{eq:reward_s1}--\eqref{eq:terminal_reward_s1} are used as the reference for reproducibility in the paper.}
\end{table*}

\subsection{Curriculum Learning: Training Environments and the Role of Success Rate}
\label{subsec:curriculum_hp_results}
Because feasibility is governed by hard operational constraints under the strict target $\theta_{\mathrm{cov}}=98\%$, we adopt success rate as the primary performance metric. An episode is counted as successful if the final deployment achieves $\theta_{\mathrm{cov}}$ while satisfying the minimum parent-connectivity requirement for all deployed non-donor nodes (i.e., no weak nodes under the stage setting). Coverage alone is insufficient in this setting, because solutions may approach the target yet remain infeasible due to connectivity violations. Success rate therefore directly quantifies the fraction of deployable solutions, while achieved coverage and the number of deployed \gls{IAB} nodes are reported as secondary indicators of solution quality and deployment efficiency.

\begin{figure*}[!t]
\centering
\includegraphics[width=\textwidth]{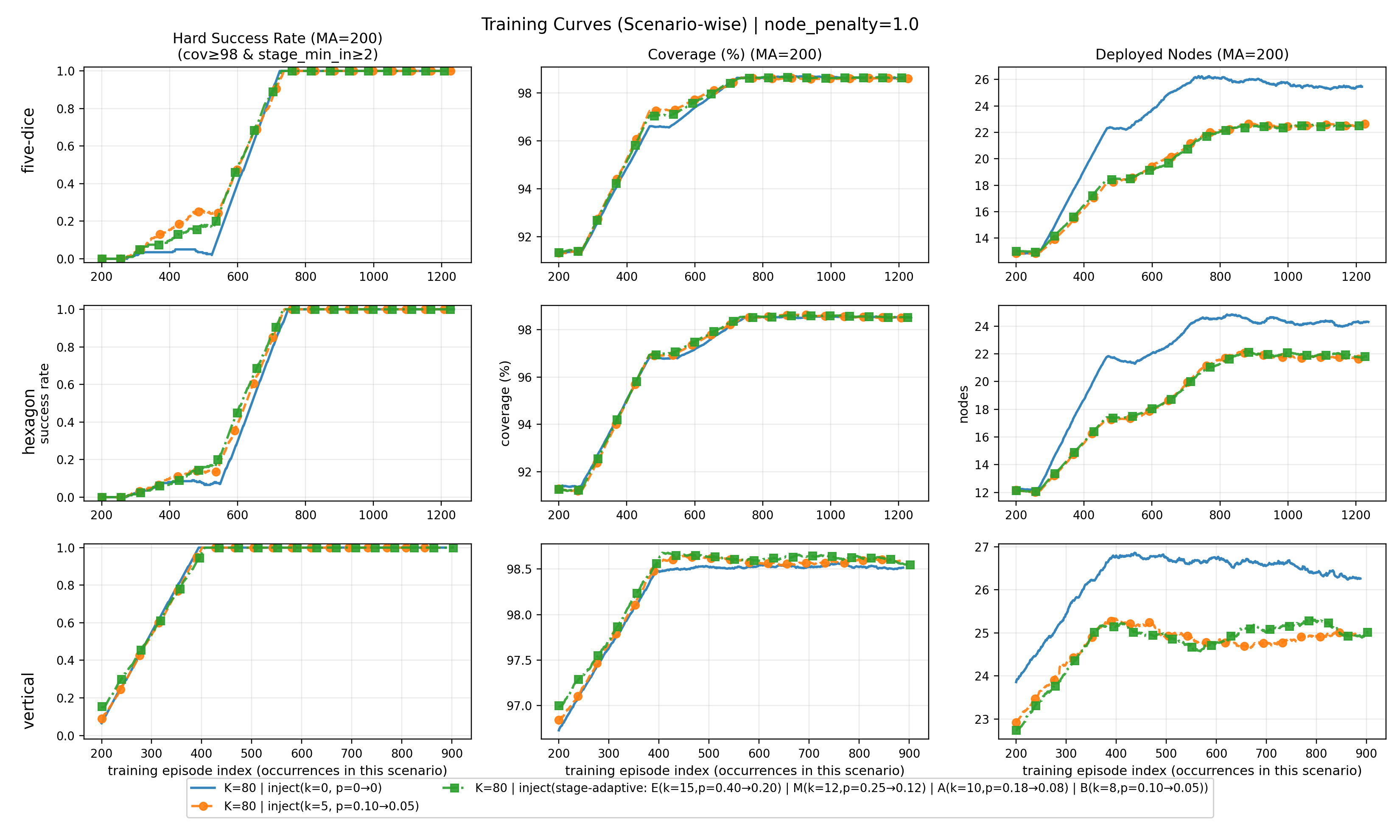}
\caption{Scenario-wise training curves (moving average $=200$) under node penalty $=1.0$ for three donor layouts (Five-Dice, Hexagon, and Vertical). Each row corresponds to one layout; the columns show hard success rate, achieved coverage, and the number of deployed \gls{IAB} nodes versus the scenario-specific training episode index.}
\label{fig:train_curves}
\end{figure*}

Fig.~\ref{fig:train_curves} reports scenario-wise training curves (moving average $=200$) and illustrates how curriculum learning improves constraint satisfaction while interacting with action-space control. Across representative layouts (Five-Dice, Hexagon, and Vertical), the hard success rate increases from near zero to approximately $1.0$ as training progresses, while achieved coverage converges to the target range. Under a node penalty ($\lambda_{\mathrm{node}}=1.0$), mild or stage-adaptive random action injection accelerates the transition to sustained hard success and reduces the number of deployed \gls{IAB} nodes at convergence without sacrificing final coverage. This behavior is consistent with the curriculum objective: once feasibility becomes learnable through staged exposure, exploratory diversity within the feasible set helps identify more deployment-efficient feasible solutions rather than merely reaching the coverage threshold.

\begin{figure}[!t]
\centering
\includegraphics[width=\columnwidth]{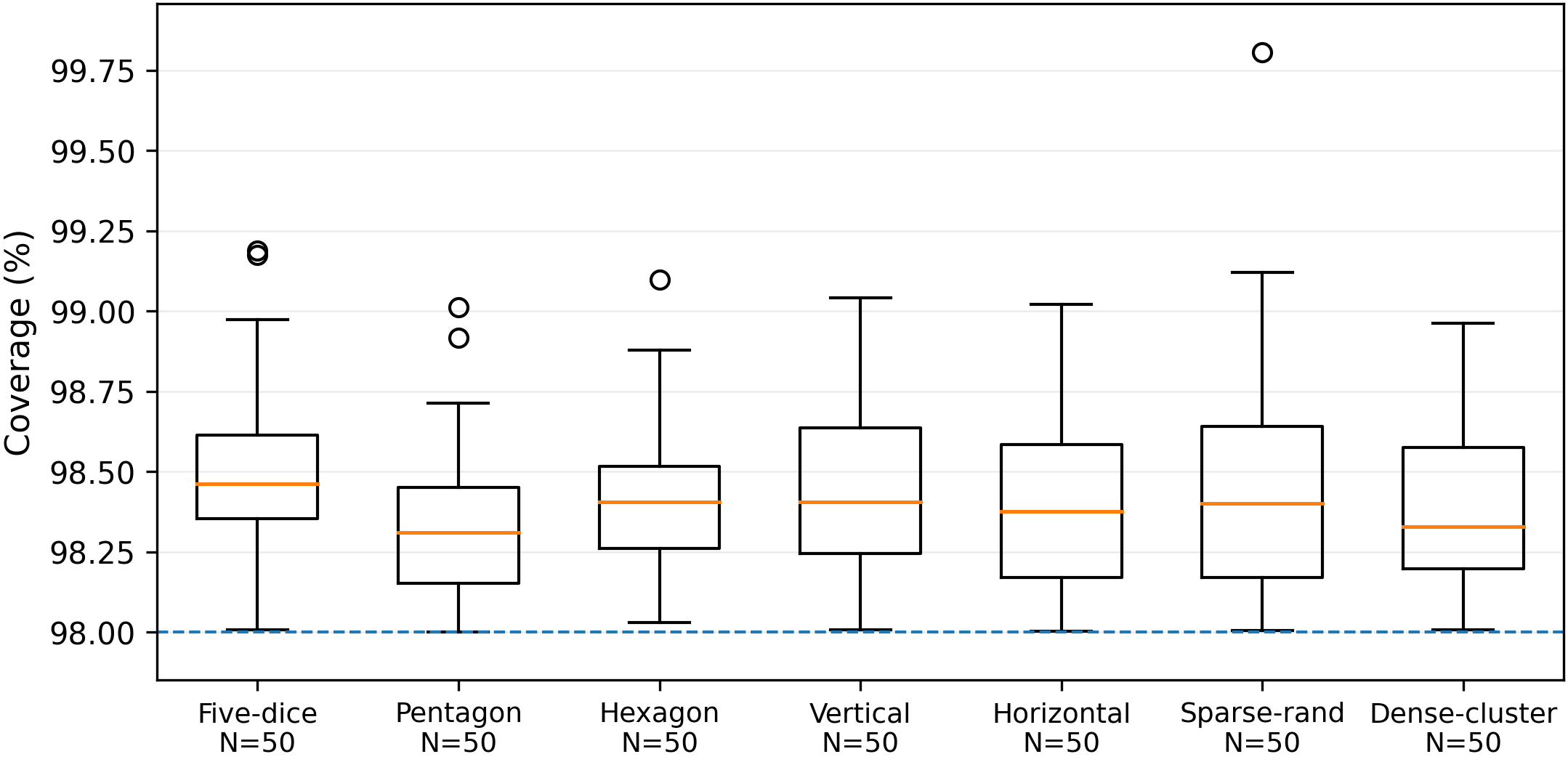}
\caption{Distribution of achieved coverage across donor scenarios over $N=50$ trials. The dashed line indicates the target threshold $\theta_{\mathrm{cov}}=98\%$.}
\label{fig:coverage_boxplot}
\end{figure}

\begin{figure}[!t]
\centering
\includegraphics[width=\columnwidth]{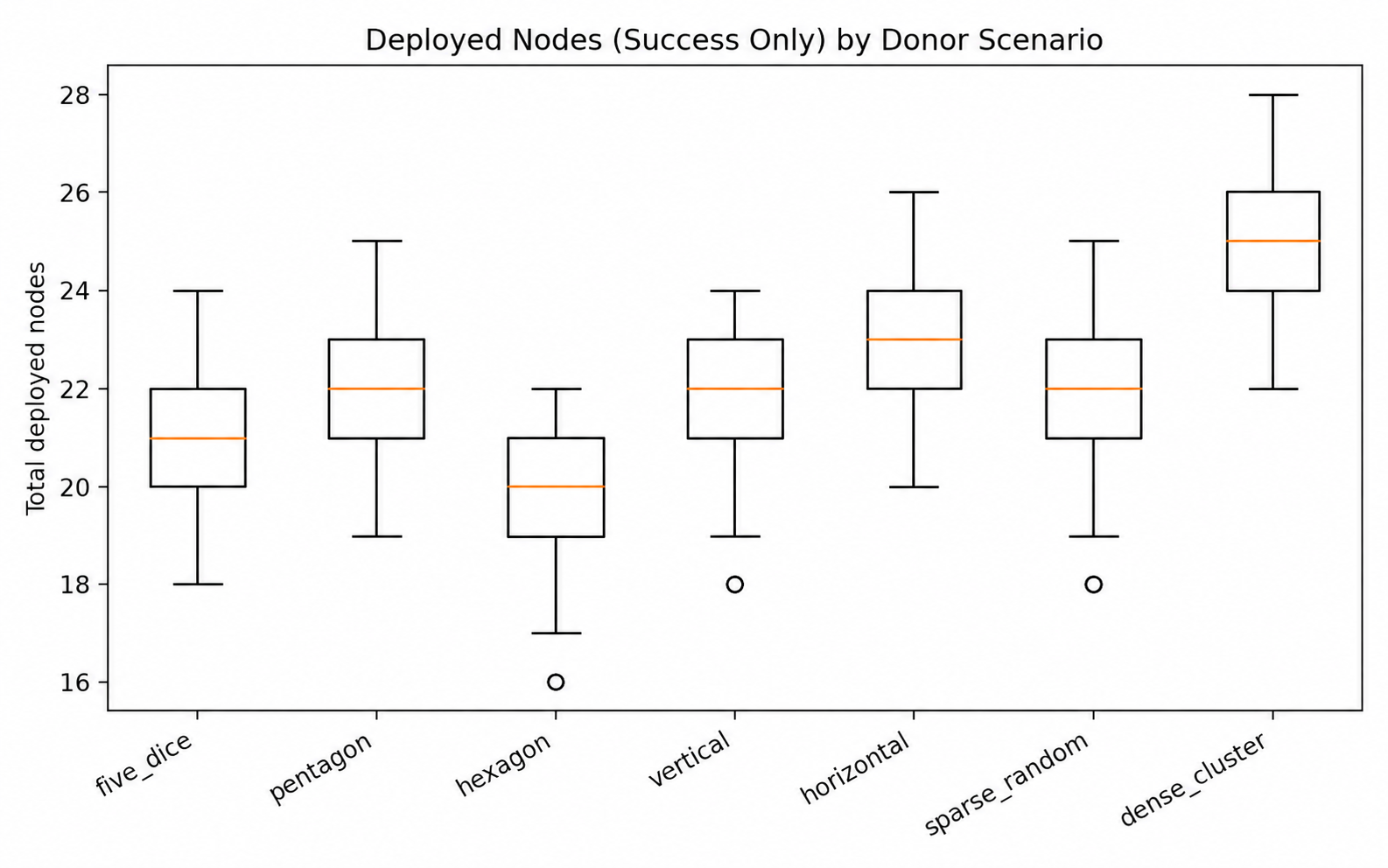}
\caption{Number of deployed \gls{IAB} nodes for successful episodes across diverse donor scenarios (median, interquartile range, whiskers, and outliers).}
\label{fig:nodes_boxplot}
\end{figure}

Crucially, curriculum learning also enables planning under random donor placements. In Sparse-Random, donor availability is inherently uncertain and varies across seeds, which changes both the initial coverage footprint and the set of feasible expansions under $m=2$. By introducing Sparse-Random only in the Hard stage, the agent is forced to adapt to this irregularity after acquiring stable, feasibility-aware deployment heuristics in regular layouts. Using the best policy selected by the unified fixed-evaluation protocol, we evaluate $N=50$ independent trials per scenario and summarize the outcome distributions in Fig.~\ref{fig:coverage_boxplot} and Fig.~\ref{fig:nodes_boxplot}. The coverage boxplot shows that the policy consistently produces solutions above the $\theta_{\mathrm{cov}}=98\%$ threshold across all donor scenarios, including Sparse-Random, with tight interquartile ranges and only occasional high-end outliers. Meanwhile, the node-count boxplot (conditioned on feasible episodes) indicates that the method maintains a controlled infrastructure cost and that the required node budget varies systematically with topology difficulty: regular and dispersed layouts concentrate around lower medians, anisotropic layouts occupy an intermediate regime, and clustered donors require more nodes. Notably, Sparse-Random exhibits stable deployment efficiency with limited dispersion across seeds, suggesting that the capability does not depend on a small set of favorable instances; instead, it emerges from staged exposure to increasing feasibility complexity, yielding a policy that transfers from regular initial conditions to irregular, seed-dependent donor placements.

\subsection{Deployment Efficiency Across Diverse
IAB Donor Topologies}
\label{subsec:efficiency}

We compare curriculum-trained GATv2--PPO with PPO
using Direct-Hard training, Graph-DQN, Dueling DQN,
and Greedy under the final planning requirements
of $\theta_{\mathrm{cov}}=98\%$ and $m=2$.
Both the proposed configuration and Direct-Hard use
PPO, but their training-task sequences differ.
Curriculum-trained GATv2--PPO progressively increases
the coverage target, the minimum number of planned
parent connections, and the diversity of IAB donor
layouts. Direct-Hard instead trains on the final
Hard-stage task from the outset, without this
staged progression.

For each evaluated IAB donor layout, each trained
learning-based algorithm generates 100 stochastic
deployment candidates. Among candidates satisfying
the static planning success criterion, the deployment
with the fewest IAB nodes is selected. Ties are
resolved by higher spatial coverage and then fewer
decision steps. Greedy provides one deterministic
deployment for each scenario. The reported outcomes
therefore describe selected deployment solutions,
rather than average performance over individual
rollouts.

\begin{figure}[!t]
\centering
\includegraphics[width=\columnwidth]{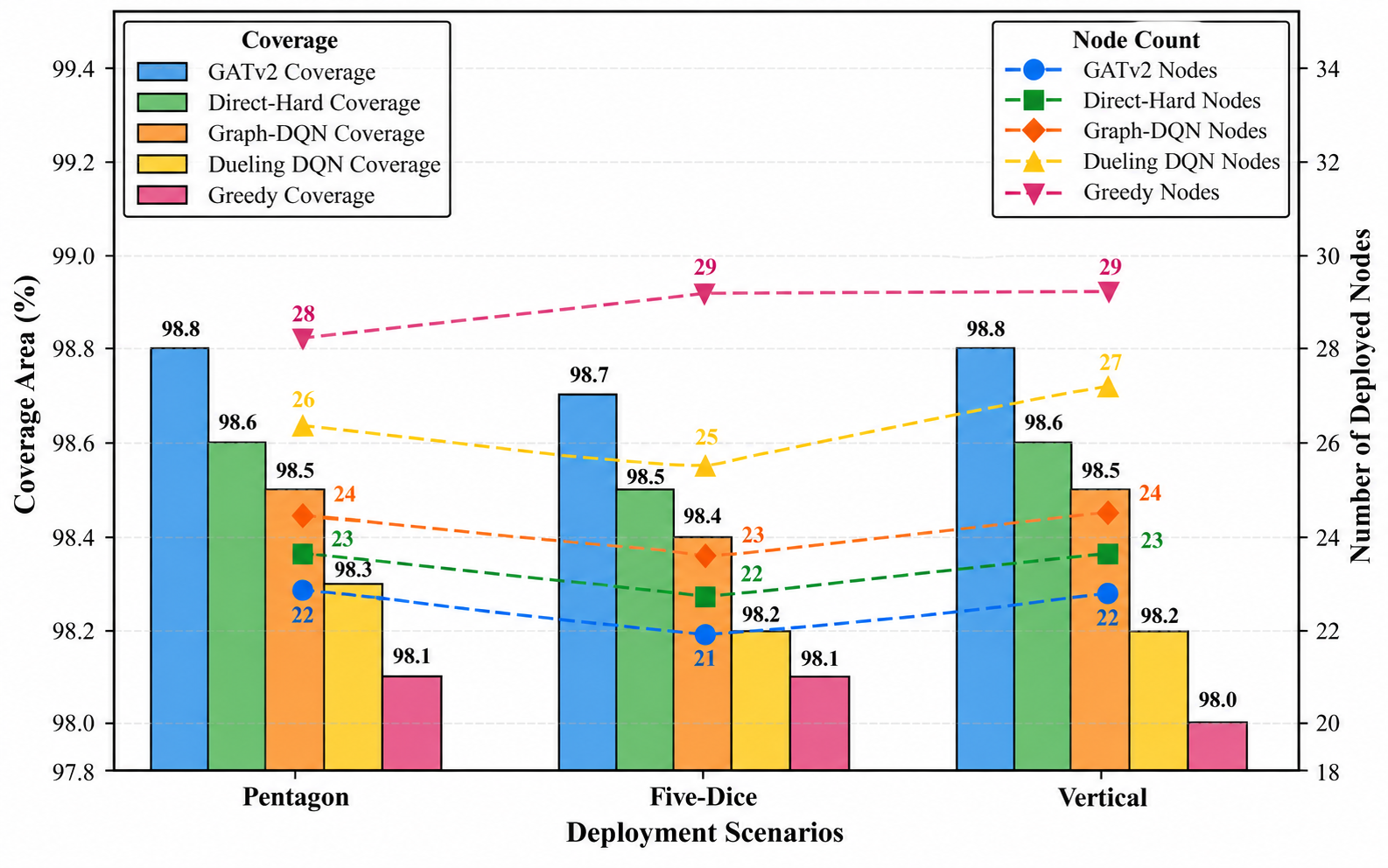}
\caption{Deployment comparison across Pentagon,
Five-Dice, and Vertical IAB donor layouts. Bars show
spatial coverage, and lines show the number of deployed
IAB nodes, excluding existing IAB donors. Learning-based
results are selected from 100 stochastic deployment
candidates; Greedy provides a deterministic reference.
Curriculum and Direct-Hard denote staged training and
direct training on the final planning task, respectively.}
\label{fig:multi_scenario}
\end{figure}

Fig.~\ref{fig:multi_scenario} shows that
curriculum-trained GATv2--PPO achieves the highest
spatial coverage and the fewest deployed IAB nodes
among the five evaluated algorithms in each of the
three reported scenarios. It achieves 98.8\% coverage
with 22 IAB nodes in Pentagon, 98.7\% with 21 IAB nodes
in Five-Dice, and 98.8\% with 22 IAB nodes in Vertical.

Compared with Direct-Hard, curriculum-trained
GATv2--PPO uses one fewer IAB node in each scenario
while achieving 0.2 percentage points higher coverage.
Compared with Graph-DQN, it uses two fewer IAB nodes
while achieving 0.3 percentage points higher coverage.
Its selected deployments also use fewer IAB nodes
than Dueling DQN and Greedy. Thus, the observed
reduction in the number of deployed IAB nodes is
not obtained by sacrificing spatial coverage.

The comparison with Direct-Hard is consistent with
the intended benefit of learning deployment decisions
under progressively stricter planning requirements,
rather than imposing the final task throughout
training. Nevertheless, these results compare
complete training configurations and do not
independently isolate every architectural and
optimization choice. The observed advantage
is restricted to the reported scenarios and
deployment-generation protocol.

\subsection{Traffic-Service Performance under Time-Varying Demand}
\label{subsec:dynamic_traffic}
The preceding experiments assess the ability of
a planning policy to determine IAB node locations and planned backhaul connections under real time-varying requirements. These infrastructure
decisions remain fixed during network operation, whereas regional demand changes with user activity and spatial concentration. The operational question is therefore how much changing demand the resulting deployment can support,
without changing its installed IAB nodes or retraining the planning policy. Daily peak--off-peak patterns and location-dependent traffic concentrations motivate this
assessment \cite{wang2024bsSleeping,commercial5GTraffic2026}.
We therefore further evaluate the fixed deployments under
a common 24-hour synthetic traffic profile. Four stress
levels are defined relative to a reference aggregate demand
of 2596.83~Mbps: low ($0.25\times$, 649.21~Mbps),
medium ($0.55\times$, 1428.26~Mbps), high
($0.85\times$, 2207.30~Mbps), and burst
($1.15\times$, 2986.35~Mbps). These scaling factors are
simulation stress levels rather than standardized cellular
load thresholds. Low, medium, high, and burst traffic are
applied during 00:00--06:00,
06:00--16:00/21:00--24:00,
16:00--18:00/19:00--21:00, and 18:00--19:00,
respectively. Spatial traffic demand additionally includes
office, residential, and event-oriented hotspots.

The evaluation uses previously trained models covers five algorithms, Dense-Cluster and Sparse-Random IAB donor topologies, and three topology realizations for each topology type. Each deployment is evaluated over all
24 hourly intervals, resulting in 720 operating points.
The same traffic-demand matrix is applied to all algorithms.
The deployed IAB nodes and planned backhaul links remain
fixed throughout the evaluation, and neither policy
retraining nor IAB node redeployment is performed.

For each deployed IAB node, one primary and one backup
upstream IAB node or IAB donor are fixed before traffic
allocation. During normal operation, at most one upstream
connection carries backhaul traffic for each IAB node.
The backup connection is activated only if the primary
backhaul connection or its upstream path to an IAB donor
becomes unavailable, rather than in response to
instantaneous traffic load. Thus, multiple planned parent
connections provide backhaul redundancy rather than
parallel traffic aggregation. The operational model further
accounts for shared access--backhaul airtime, half-duplex
operation, IAB donor fiber-capacity constraints, and a
400~MHz backhaul bandwidth. The link budget includes
oxygen absorption, rain attenuation, RF loss, fading
margin, and a fixed interference margin.

For each hourly operating point, the served-demand ratio
is defined as
\begin{equation}
\eta_{\mathrm{srv}}
=
\frac{\displaystyle\sum_{k\in\mathcal K_d}s_k}
     {\displaystyle\sum_{k\in\mathcal K_d}d_k},
\label{eq:served_demand_ratio}
\end{equation}
where $d_k$ and $s_k$ denote the offered and served traffic
rates, respectively, at demand-grid location $k$. The
denominator includes all offered traffic in the evaluation
area, including traffic that cannot be served. Hence,
$\eta_{\mathrm{srv}}$ measures operational traffic support
and is distinct from the spatial coverage ratio used for
deployment planning.

Table~\ref{tab:dynamic_single_parent} focuses on the
high-load and burst-load conditions, which provide the
strongest traffic stress on the fixed deployments.
Individual comparison algorithms achieve the highest
served-demand ratio under some specific topology--load
conditions. Nevertheless, curriculum-trained GATv2--PPO
provides the strongest overall cross-condition performance.
Its descriptive average over the four reported
topology--load conditions is 82.18\%, compared with
80.59\% for PPO with Direct-Hard training, 80.30\% for
Dueling DQN, 73.73\% for Graph-DQN, and 67.37\% for
Greedy.

To further quantify consistency across heterogeneous
operating conditions, we calculate the mean gap from the
best-performing algorithm in each topology--load condition.
Curriculum-trained GATv2--PPO has the smallest mean gap,
at only 4.04 percentage points, compared with 5.63 for
PPO with Direct-Hard training, 5.92 for Dueling DQN,
12.49 for Graph-DQN, and 18.85 for Greedy. Therefore,
although curriculum-trained GATv2--PPO is not the
highest-performing algorithm in every individual operating
condition, it remains closest to the condition-specific
best performance on average.

The advantage is particularly pronounced under the
Dense-Cluster high-load condition. Curriculum-trained
GATv2--PPO serves 92.83\% of the offered traffic,
exceeding PPO with Direct-Hard training by 7.26 percentage
points, Dueling DQN by 2.79 percentage points, Graph-DQN
by 28.41 percentage points, and Greedy by 20.49 percentage
points. Under Sparse-Random burst traffic,
curriculum-trained GATv2--PPO achieves 85.43\%, effectively
matching the highest value of 85.50\% obtained by
Graph-DQN, while exceeding PPO with Direct-Hard training,
Dueling DQN, and Greedy by 10.03, 8.66, and
16.63 percentage points, respectively.

These results indicate that curriculum-trained GATv2--PPO
provides a more balanced response to simultaneous changes
in IAB donor topology and traffic stress. This behavior is
consistent with the curriculum design: the policy is
progressively exposed to increasingly stringent coverage
and backhaul-connectivity requirements and increasingly
irregular IAB donor topologies. Such staged exposure
encourages deployment decisions that remain effective
under different multi-hop expansion conditions rather than
specializing to a single IAB donor topology. Importantly,
the dynamic traffic profiles are introduced only after
deployment and are not used for policy retraining.
Therefore, the results demonstrate robustness of the
resulting fixed deployments rather than online
traffic-aware adaptation by GATv2--PPO.

The cross-condition average and mean gap are descriptive
summary metrics that assign equal weight to the four
reported stress conditions; they are not statistical
significance tests. Random time-varying blockage and
interference, UE mobility, packet-level queuing, and
online GATv2--PPO adaptation remain outside the scope of
the present evaluation.

\begin{table*}[!t]
\centering
\caption{Traffic-service performance under time-varying demand resilience operation.
The cross-condition average is
the arithmetic mean of the four reported median
served-demand ratios. The mean gap to condition-best is the average percentage-point difference from the highest served-demand ratio in each condition; lower is better.}
\label{tab:dynamic_single_parent}
\footnotesize
\setlength{\tabcolsep}{3.6pt}
\renewcommand{\arraystretch}{1.08}
\begin{tabular*}{\textwidth}
{@{\extracolsep{\fill}}lcccccc@{}}
\toprule
&
\multicolumn{2}{c}{\textbf{Dense-Cluster}} &
\multicolumn{2}{c}{\textbf{Sparse-Random}} &
\shortstack{\textbf{Cross-condition}\\\textbf{average}} &
\shortstack{\textbf{Mean gap to}\\\textbf{condition-best}} \\
\cmidrule(lr){2-3}
\cmidrule(lr){4-5}

\textbf{Algorithm} &
\shortstack{\textbf{High}\\\textbf{(\%)}} &
\shortstack{\textbf{Burst}\\\textbf{(\%)}} &
\shortstack{\textbf{High}\\\textbf{(\%)}} &
\shortstack{\textbf{Burst}\\\textbf{(\%)}} &
\textbf{(\%)} &
\textbf{(pp)} \\
\midrule

\textbf{GATv2--PPO (Curriculum)}
& \textbf{92.83}
& 64.46
& 86.00
& 85.43
& \textbf{82.18}
& \textbf{4.04} \\

PPO (Direct-Hard)
& 85.57
& \textbf{73.38}
& 88.01
& 75.40
& 80.59
& 5.63 \\

Graph-DQN
& 64.42
& 54.14
& 90.86
& \textbf{85.50}
& 73.73
& 12.49 \\

Dueling DQN
& 90.04
& 61.24
& \textbf{93.15}
& 76.77
& 80.30
& 5.92 \\

Greedy
& 72.34
& 51.64
& 76.70
& 68.80
& 67.37
& 18.85 \\

\bottomrule
\end{tabular*}
\end{table*}
\subsection{Resilience Under Progressive Backhaul Failures}
\label{subsec:resilience}
We evaluate resilience using progressive backhaul-failure tests that emulate \gls{mmWave} blockage and abrupt link degradation. For each scenario instance, we first construct a feasible deployment using the same best-selection protocol described above for GATv2--PPO and the baselines. Starting from this fixed deployment, we randomly disable a fraction of active backhaul links (10\%, 20\%, and 30\%) and recompute the maintained coverage, defined as the fraction of grid cells that still satisfy the coverage criterion under the remaining connectivity. No replanning is performed after failures, so the evaluation isolates the intrinsic robustness of the constructed backhaul topology. For each failure ratio, random link disabling is repeated across multiple realizations, and we report the average maintained coverage.

\begin{figure}[!t]
\centering
\includegraphics[width=\columnwidth]{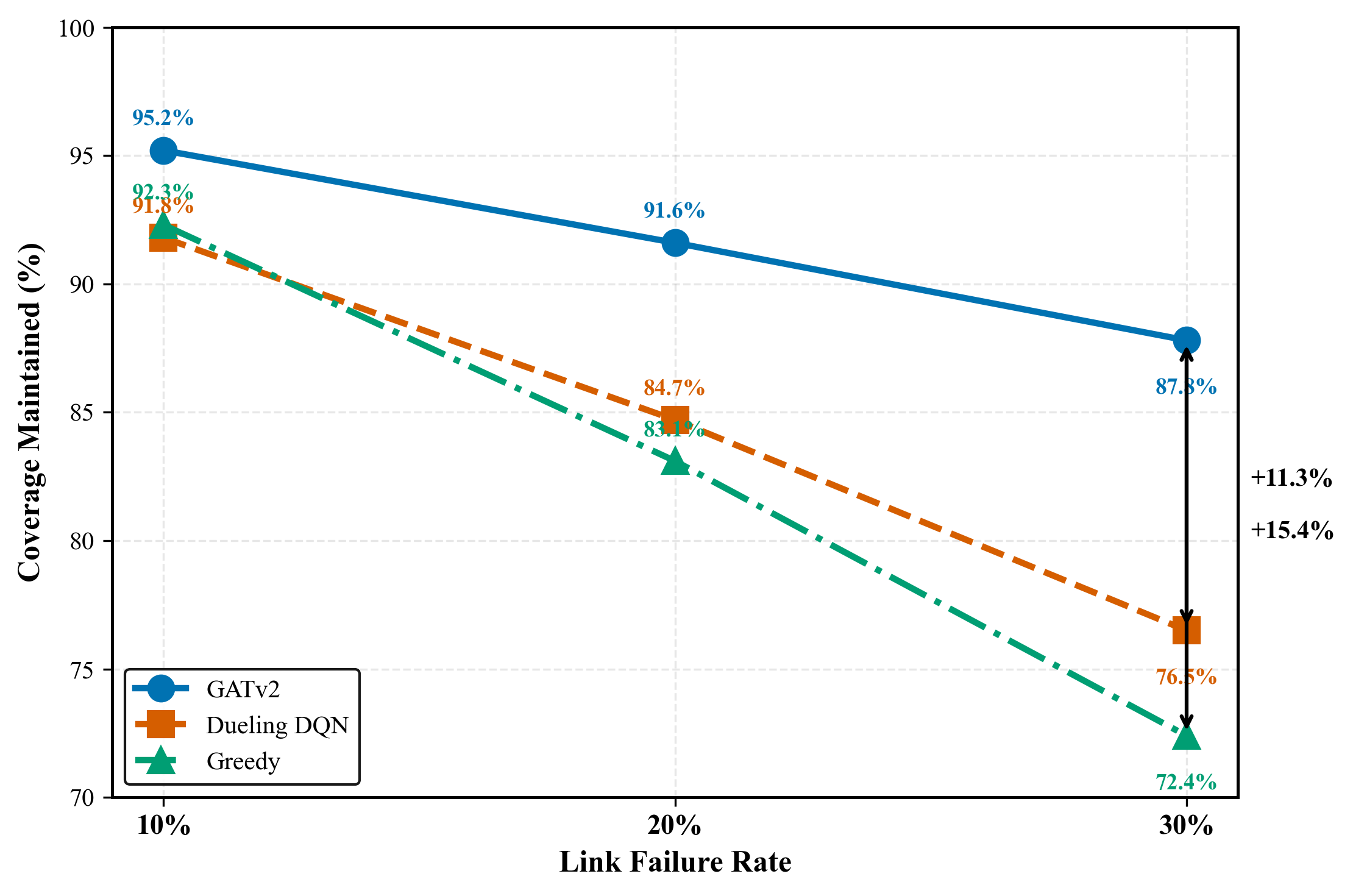}
\caption{Progressive link-failure evaluation: maintained coverage after randomly disabling 10\%, 20\%, and 30\% of active backhaul links.}
\label{fig:resilience_analysis}
\end{figure}

As shown in Fig.~\ref{fig:resilience_analysis}, GATv2--PPO maintains higher post-failure coverage than both baselines at all failure levels: 95.2\% at 10\% failures, 91.6\% at 20\% failures, and 87.8\% at 30\% failures. In contrast, Dueling DQN decreases to 76.5\% and the greedy baseline decreases to 72.4\% under 30\% failures, indicating substantially greater degradation under severe link disruptions. This advantage is consistent with enforcing $m=2$ parent-connectivity during deployment construction, which promotes redundancy in the backhaul structure and reduces reliance on a small set of critical links.

\subsection{Computational Complexity and Scalability}
\label{subsec:complexity_scalability}

Let $N_{\mathrm c}=|\mathcal J|$ denote the number of
candidate IAB node locations, and let
$N_{\mathrm v}=|\mathcal V|$ and
$N_{\mathrm e}=|\mathcal E|$ denote the numbers of
graph vertices and candidate backhaul links.
For $L$ graph-encoding layers, hidden dimensions bounded
by $d$, and fixed attention-head and edge-feature
dimensions, sparse GATv2 evaluation and shared
action scoring require
\begin{equation}
\mathcal C_{\mathrm{pol}}
=
O\!\left(
L\left[N_{\mathrm v}d^2+N_{\mathrm e}d\right]
\right).
\label{eq:policy_complexity}
\end{equation}
If $|\mathcal I|=O(N_{\mathrm c})$, then
$N_{\mathrm v}=O(N_{\mathrm c})$.
For fixed network dimensions, the per-step policy
computation is therefore $O(N_{\mathrm c})$ when
the average graph degree remains bounded, but can
be $O(N_{\mathrm c}^2)$ for dense candidate graphs.
Increasing candidate-location density can increase
both the number of vertices and the average degree.

Generating $N_{\mathrm{cand}}$ deployment candidates
requires
\begin{equation}
\begin{aligned}
\mathcal C_{\mathrm{plan}}
={}&\mathcal C_{\mathrm{init}}
+\mathcal C_{\mathrm{select}}\\
&+\sum_{b=1}^{N_{\mathrm{cand}}}
\sum_{t=0}^{T_b-1}
\left(
\mathcal C_{\mathrm{pol}}
+\mathcal C_{\mathrm{env},b,t}
\right),
\end{aligned}
\label{eq:planning_complexity}
\end{equation}
where $T_b$ is the decision count for candidate $b$.
Initialization includes shared preprocessing and
per-candidate resets; selection includes final
validation. Environment updates account for coverage,
connectivity, and rate-budget calculations.
For example, scanning one precomputed coverage mask
costs $O(|\mathcal K|)$, whereas sorting $q_t$ eligible
upstream IAB nodes or IAB donors costs
$O(q_t\log q_t)$. Additional rescoring and budget
propagation remain implementation-dependent.

When each deployment action selects a previously
undeployed IAB node, $T_b\leq N_{\mathrm c}+1$,
including a possible stop action. Consequently,
the cumulative policy-evaluation component has
worst-case bounds of
$O(N_{\mathrm{cand}}N_{\mathrm c}^{2})$
for bounded-degree graphs and
$O(N_{\mathrm{cand}}N_{\mathrm c}^{3})$
for dense graphs, with fixed network dimensions.
These bounds exclude the separately identified
environment and initialization costs. Explicit
coverage-mask storage additionally requires
$O(N_{\mathrm v}|\mathcal K|)$ memory.

For $S$ retained training transitions and
$K_{\mathrm{PPO}}$ optimization epochs per batch,
the offline training cost can be decomposed as
\begin{equation}
\begin{aligned}
\mathcal C_{\mathrm{train}}
={}&\mathcal C_{\mathrm{collect}}
+O\!\left(
K_{\mathrm{PPO}}S\overline{\mathcal C}_{\mathrm{pol}}
\right)\\
&+\mathcal C_{\mathrm{schedule}}
+\mathcal C_{\mathrm{validation}},
\end{aligned}
\label{eq:training_complexity}
\end{equation}
where $\overline{\mathcal C}_{\mathrm{pol}}$ represents
the average network-evaluation cost.
Curriculum scheduling operates during training and
adds no scheduling operation during deployment inference.
The previously reported 6000-episode training cost
of approximately 10 hours is separate from the
deployment-generation measurements below.

Table~\ref{tab:scalability_compact} reports the proposed
method at 900 and 1600 candidate locations.
Each setting combines three trained models and ten
shared IAB donor layouts, giving 30 evaluations.
Measurements use an NVIDIA GeForce RTX 4090 Laptop GPU.
Each case generates one deployment candidate, not
100 candidates. Mean times retain failed evaluations
and exclude separately recorded model/map loading,
warm-up, and IAB donor-list generation.

Sparse-Random achieves 30/30 successful evaluations
at both sizes, whereas Dense-Cluster decreases from
22/30 to 9/30. Thus, the method remains computationally
executable at larger candidate-set sizes, but planning
success remains dependent on the IAB donor layout.

\begin{table}[!t]
\centering
\caption{Scale evaluation of curriculum-trained
GATv2--PPO. Times are arithmetic means over all
evaluations, including failures; success is reported
as successful/total evaluations.}
\label{tab:scalability_compact}
\footnotesize
\setlength{\tabcolsep}{3pt}
\renewcommand{\arraystretch}{1.1}
\begin{tabularx}{\columnwidth}{
@{}r >{\raggedright\arraybackslash}X cc@{}}
\toprule
\shortstack{\textbf{Candidate}\\\textbf{locations}} &
\textbf{IAB donor layout} &
\shortstack{\textbf{Mean time}\\\textbf{(s)}} &
\textbf{Success} \\
\midrule
900  & Sparse-Random & 0.946 & 30/30 \\
900  & Dense-Cluster & 1.444 & 22/30 \\
1600 & Sparse-Random & 2.264 & 30/30 \\
1600 & Dense-Cluster & 5.031 & 9/30 \\
\bottomrule
\end{tabularx}
\end{table}
\section{Conclusion}
\label{sec:conclusion}

Deploying mmWave \gls{IAB} networks for 6G infrastructures requires planning mechanisms that can generalize across heterogeneous donor topologies while enforcing stringent coverage and resilience requirements under multi-hop transport constraints. Existing approaches involve fundamental trade-offs: optimization-based methods offer strong feasibility guarantees but scale poorly, heuristics often lack systematic resilience modeling, and learning-based approaches can suffer from limited generalization when donor configurations change. This work addresses these limitations by introducing a graph neural curriculum learning framework that combines explicit constraint handling with progressive task decomposition.

Our contributions are threefold. First, we formulate resilient \gls{IAB} deployment as a constraint-aware \gls{MDP} and represent the evolving network as an attributed graph, enabling a \gls{GATv2}-based actor--critic to capture donor-induced multi-hop dependencies via edge-conditioned attention, with action masking to enforce feasibility. Second, we develop a four-stage curriculum learning strategy with progressive difficulty scaling (coverage 90\%$\rightarrow$98\%, resilience $m=1\rightarrow 2$, and topology regular$\rightarrow$irregular) and stability mechanisms (disaster isolation, soft mixing, and unified evaluation) that mitigate catastrophic forgetting and improve convergence under sparse feasibility signals. Third, the reported evaluations support deployment performance across the tested IAB donor layouts without retraining.

Experiments across seven diverse \gls{IAB} donor scenarios show consistent gains. The proposed method achieves the 98\% coverage target under dual-connectivity constraints with 8.33--27.59\% fewer deployed nodes than baseline methods, maintains high feasibility across all topologies---including Sparse-Random---with stable deployment outcomes, and improves robustness under progressive backhaul failures, sustaining 87.8\% coverage at 30\% link failures compared with 72--76\% for the baselines. The full curriculum can be trained in approximately 10 hours on a single GPU over 6000 episodes.

The practical contribution is an offline planner that combines coverage, connectivity, and deployment cost before assessing fixed deployments under changing demand and link availability. Spatial coverage, served demand, and post-failure coverage answer distinct questions and are reported separately. Computational execution at larger candidate sets does not guarantee feasible planning for every IAB donor layout: Dense-Cluster success falls from 22/30 at 900 candidate locations to 9/30 at 1600. Future work will investigate donor-layout-aware training and calibration using deployment measurements to address these planning and model-fidelity limitations.

\bibliographystyle{IEEEtran}

\bibliography{refs}

\vfill

\end{document}